\documentclass[preprint,aps, amsmath,amssymb]{revtex4-1}
\newcommand{\be}{\begin{equation}}
\newcommand{\ee}{\end{equation}}
\newcommand{\bea}{\begin{eqnarray}}
\newcommand{\eea}{\end{eqnarray}}
\newcommand{\ba}{\begin{array}}
\newcommand{\ea}{\end{array}}

\usepackage[utf8]{inputenc}
\usepackage{soulutf8}
\usepackage{graphicx}
\usepackage{epstopdf}
\usepackage{amsmath}
\usepackage{slashed}
\usepackage{mathtools}
\usepackage{slashed}
\usepackage{array,multirow}

\begin{document}

\title{Predicting the light-front holographic TMDs of the pion}

\author{Mohammad Ahmady}
\email{mahmady@mta.ca}
\affiliation{\small Department of Physics, Mount Allison University, \mbox{Sackville, New Brunswick, Canada, E4L 1E6}}

\author{Chandan Mondal}
\email{mondal@impcas.ac.cn}
\affiliation{\small Institute of Modern Physics, Chinese Academy of Sciences, \mbox{Lanzhou-730000, China}}

\author{Ruben Sandapen}
\email{ruben.sandapen@acadiau.ca}
\affiliation{\small Department of Physics, Acadia University, \mbox{Wolfville, Nova-Scotia, Canada, B4P 2R6}}

\begin{abstract} 
We predict the twist-2 Transverse Momentum Dependent parton distribution functions (TMDs) of the pion, namely the unpolarized quark TMD, $f_{1}(x, k_\perp)$, and the transversely polarized quark TMD, also known as the Boer-Mulders function, $h^\perp_{1}(x, k_\perp)$, using a holographic light-front pion wavefunction with dynamical spin effects. These spin effects, in conjunction with gluon rescattering, are crucial to predict a non-zero holographic Boer-Mulders function. We investigate the use of a non-perturbative SU(3) gluon rescattering kernel, thus going beyond the usual approximation of perturbative U(1) gluons. We also predict the generalized Boer-Mulders shifts in order to compare with the available lattice data. 

\end{abstract}

\maketitle
         
\section{Introduction}

\label{Sec:Introduction}

Transverse Momentum Dependent parton distributions functions (TMDs) contain important information on the three-dimensional internal structure of hadrons, especially the spin-orbit correlations of quarks within them \cite{Angeles-Martinez:2015sea}. For the pion, there are two twist-$2$ TMDs: the unpolarized quark TMD, $f_1(x, k_\perp)$, and the polarized quark TMD, $h_1^{\perp}(x, k_\perp)$, also known as the Boer-Mulders function \cite{Boer:1997nt,Boer:1999mm}. $f_1(x,k_\perp)$  describes the momentum distribution of unpolarized quarks within the pion while  $h_1^{\perp}(x, k_\perp)$ describes the spin-orbit correlations of transversely polarized quarks within the pion.  The Boer-Mulders function is naively a T-odd distribution and such distributions were initially thought to vanish due to the time reversal invariance of QCD \cite{Collins:1992kk} but later it became apparent that they can be dynamically generated by initial or final state interactions \cite{Brodsky:2002cx, Brodsky:2002rv}. More formally, T-odd distributions do not vanish due to the non-trivial gauge link that guarantees the colour gauge invariance of their field-theoretic definitions \cite{Collins:2002kn,Ji:2002aa,Belitsky:2002sm}. At the same time, the gauge link makes these distributions process-dependent, flipping their sign from Semi Inclusive Deep Inelastic Scattering (SIDIS) to Drell-Yan (DY) scattering. 

In conjunction with their nucleon counterparts, the pion unpolarized TMD and the Boer-Mulders function are inputs in the theoretical predictions of the cross-sections and azimuthal asymmetries for unpolarized pion-induced DY scattering \cite{Pasquini:2014ppa,Wang:2017onm} which have both been measured \cite{Guanziroli:1987rp,Falciano:1986wk,Conway:1989fs}. The azimuthal asymmetry has been observed to be large and a better theoretical understanding of the pion Boer-Mulders function may help to explain this observation. Otherwise, little is known from experiment on the pion TMDs although this situation is likely to change with the new COMPASS collaboration program of pion-induced DY scattering \cite{Aghasyan:2017jop,COMPASS}. 

On the theory side, the pion Boer-Mulders function has been predicted in the antiquark spectator model \cite{Lu:2004au,Meissner:2008ay}, in the light-front constituent quark model \cite{Wang:2017onm,Wang:2018naw,Pasquini:2014ppa,Lorce:2016ugb}, in the MIT bag model \cite{Lu:2012hh} and in the  Nambu-Jona-Lasinio model \cite{Noguera:2015iia,Ceccopieri:2018nop}. In all cases, the physical mechanism generating the Boer-Mulders function is perturbative gluon rescattering. A notable attempt to go beyond this perturbative approximation has been made by Gamberg and Schlegel \cite{Gamberg:2009uk} within the antiquark spectator framework. The pion TMDs have also been studied on the lattice \cite{Brommel:2007xd,Engelhardt:2015xja}. 

Our goal in this paper is to predict the pion TMDs using the spin-improved holographic pion light-front wavefunction \cite{Ahmady:2016ufq,Ahmady:2018muv}. We shall generate the holographic Boer-Mulders using the non-perturbative kernel of Ref. \cite{Gamberg:2009uk} as well as its perturbative limit. Finally, we shall predict the so-called generalized Boer-Mulders shifts in order to compare with the lattice data of Ref. \cite{Engelhardt:2015xja}.

\section{Holographic light-front wavefunctions}

The holographic pion wavefunction is obtained by solving the holographic Schr\"odinger equation for mesons \cite{deTeramond:2005su,Brodsky:2006uqa,deTeramond:2008ht}: 
 \begin{equation}
			\left(-\frac{\mathrm{d}^2}{\mathrm{d}\zeta^2}-\frac{1-4L^2}{4\zeta^2} + U_{\mathrm{eff}}(\zeta) \right) \phi(\zeta)=M^2 \phi(\zeta) \;,
	\label{hSE}
	\end{equation}
	with
\begin{equation}
	\mathbf{\zeta} = \sqrt{x\bar{x}} b \hspace{1cm} \hspace{1cm}  \hspace{1cm} (\bar{x} \equiv 1-x	)\;,
	\end{equation}
where $b$ is the transverse separation of the quark and antiquark and $x$ is the light-front momentum fraction carried by the quark, 
and
	\begin{equation}
		U_{\mathrm{eff}}(\zeta)=\kappa^4 \zeta^2 + 2 \kappa^2 (J-1)	\;,
	\label{hUeff}	
	\end{equation}
 where $J=L+S$. Solving Eq. \eqref{hSE} yields the meson mass spectrum 
\begin{equation}
 	M^2= 4\kappa^2 \left(n+L +\frac{S}{2}\right) \;,
 	\label{mass-Regge}
 \end{equation}
and the wavefunctions
 \begin{equation}
 	\phi_{nL}(\zeta)= \kappa^{1+L} \sqrt{\frac{2 n !}{(n+L)!}} \zeta^{1/2+L} \exp{\left(\frac{-\kappa^2 \zeta^2}{2}\right)}  ~ L_n^L(\kappa^2 \zeta^2)\;.
 \label{phi-zeta}
 \end{equation}
As can be seen from Eq. \eqref{mass-Regge}, the lightest bound state, with $n=S=L=0$, is massless and is naturally identified with the pion. In holographic light-front QCD, the massless pion is a consequence of the unique form of the holographic confining potential given by Eq. \eqref{mass-Regge}. The harmonic term, $\kappa^4 \zeta^2$, of the confining potential is obtained by the de Alfaro, Furbini and Furlan (dAFF) \cite{deAlfaro:1976vlx} mechanism which enables a mass scale $\kappa$ to appear in the equations of motion while conformal symmetry is still preserved in the underlying action. Then, the spin term, $2\kappa^2(J-1)$, results from the holographic mapping between massless light-front QCD and a string theory in the higher dimensional anti-de Sitter (AdS) space. In this gauge/gravity duality, the radial light-front variable $\zeta$ maps onto the fifth dimension in AdS space and the mass scale $\kappa$ emerging from the dAFF mechanism governs simultaneously the strength of the confining harmonic potential in physical spacetime and that of the dilaton field which breaks conformal invariance in AdS space. Hence, we refer to $\kappa$ as the AdS/QCD mass scale. As can be seen from Eq. \eqref{mass-Regge}, the AdS/QCD scale $\kappa$ fixes the slope of the experimentally observed Regge trajectories and can thus be extracted from spectroscopic data. A simultaneous fit to the Regge slopes of mesons and baryons  gives $\kappa=(523 \pm 24)$ MeV \cite{Brodsky:2016rvj} which we refer to as the universal AdS/QCD scale.

 The holographic light-front Schr\"odinger Equation only gives the radial part of the meson light-front wavefunction. The complete wavefunction is given by \cite{Brodsky:2014yha}
\begin{equation}
	\Psi (x,\zeta,\varphi)=\frac{\phi(\zeta)}{\sqrt{2\pi \zeta}} X(x) e^{iL\varphi} \;,
	\end{equation}
where $X(x)=\sqrt{x\bar{x}}$ as obtained by a precise mapping of the electromagnetic form factor in AdS and in physical spacetime \cite{Brodsky:2008pf}. The normalized holographic light-front wavefunction for the pion is then given by
 \begin{equation}
 	\Psi (x,\zeta^2) = \frac{\kappa}{\sqrt{\pi}} \sqrt{x \bar{x}}  \exp{ \left[ -{ \kappa^2 \zeta^2  \over 2} \right] } \;.
\label{pionhwf} 
\end{equation}
So far, the quark masses have been neglected and their spins ignored. Assuming that there is no spin-orbit correlation in the pion, it is straightforward to restore independently the dependence of the wavefunction on the quark masses and helicities: 
\begin{equation}
\Psi (x,\zeta^2) \propto  \sqrt{x \bar{x}}  \exp{ \left[ -\frac{\kappa^2 \zeta^2}{2} \right] } 
 \exp{ \left[ - {m_{f}^2 \over 2 \kappa^2 x \bar{x} } \right]} \frac{1}{\sqrt{2}} h\delta_{h,-\bar{h}} \;.
\label{pion-hwf-quark-masses}
\end{equation}
  
We have shown in previous papers \cite{Ahmady:2016ufq,Ahmady:2018muv}  that it is possible to achieve a very successful pion phenomenology by allowing for dynamical spin effects (i.e. spin-orbit correlations) in the pion. We do so by using a spin-improved holographic wavefunction given, in momentum space, by
 
\begin{equation}
	\Psi_{h \bar{h}}(x,\mathbf{k}) = \Psi(x, \mathbf{k}) S_{h \bar{h}} (x, \mathbf{k}) \;,
	\label{spin-improved-wf}
	\end{equation}
where 
\begin{equation}
	S_{h \bar{h}} (x, \mathbf{k})= \frac{\bar{u}_{h}(x,\mathbf{k})}{\sqrt{\bar{x}}} \left[\frac{M_\pi}{2P^+} \gamma^+ \gamma^5 + B  \gamma^5 \right] \frac{v_{\bar{h}}(x,\mathbf{k})}{\sqrt{x}} \;,
	\label{spin-structure} 
	\end{equation}
 and
	 \begin{equation}
	 	\Psi (x,\mathbf{k})=\mathcal{N} \frac{1}{\sqrt{x \bar{x}}}  \exp{ \left[ -\frac{k_\perp^2+m_f^2}{2\kappa^2 x\bar{x}} \right] } 
 \label{hWF-k}
 \end{equation}
 is the two-dimensional Fourier transform of the holographic pion wavefunction given by Eq. \eqref{pionhwf}. Here $k_\perp=|\mathbf{k}|$: in this paper, we use the notation $a_\perp=|\mathbf{a}|$ where $\mathbf{a}$ is any 2-dimensional momentum. $\mathcal{N}$ is a normalization constant fixed using
 \begin{equation}
 	\sum_{h,\bar{h}}\int \mathrm{d} x \frac{\mathrm{d}^2 \mathbf{k}}{16\pi^3} |\Psi_{h \bar{h}}(x,\mathbf{k})|^2 =1 \;.
\end{equation}

We refer to $B$ as the dynamical spin parameter: $B \to 0$ means no spin-orbit correlations as in the original holographic wavefunction, while, on the other hand, $B \ge 1$ corresponds to a maximal spin-orbit correlations.  The resulting spin-improved holographic wavefunction is then given by 
	\begin{eqnarray}
	 	\Psi_{h,\bar{h}}(x,\mathbf{k})= \left[ (M_{\pi} x\bar{x} + B m_f) h\delta_{h,-\bar{h}}  - B    k_\perp e^{-ih\theta_{k_\perp}}\delta_{h,\bar{h}}	\right] \frac{\Psi (x, k_\perp^2)}{x\bar{x}} 
	 \label{spin-improved-wfn-k}
	 \end{eqnarray}
which we can rewrite as
\begin{equation}
	\Psi_{h,\bar{h}}(x,\mathbf{k})= [\Psi_{h,\bar{h}}(x,\mathbf{k})]^{L_z=S_z=0} + [\Psi_{h,\bar{h}}(x,\mathbf{k})]^{L_z=-S_z} 
	\end{equation}	 
where
\begin{equation}
	[\Psi_{h,\bar{h}}(x,\mathbf{k})]^{L_z=S_z=0}=(M_{\pi} x\bar{x} + B m_f) h\delta_{h,-\bar{h}} \frac{\Psi (x, k_\perp^2)}{x\bar{x}} 	
\end{equation}	
and 
\begin{equation}
	[\Psi_{h,\bar{h}}(x,\mathbf{k})]^{L_z=-S_z}=- (B    k e^{-ih\theta_{k_\perp}}\delta_{h,\bar{h}})	\frac{\Psi (x, k_\perp^2)}{x\bar{x}} 
\end{equation}	
to highlight the fact that dynamical spin effects are accounted for by two corrections to the original holographic wavefunction: a term proportional to the quark mass (which therefore vanishes in the chiral limit) and a  new $(L_z =\pm 1, S_z=\mp 1)$ component which allows for the spins of the quarks to be aligned and which actually survives in the chiral limit.  We shall see that the latter is directly responsible for a non-zero Boer-Mulders function.

With $B \ge 1$, $m_{u/d}=330$ MeV and $\kappa=523$ MeV, we successfully predict simultaneously the pion decay constant, charge radius, EM and transition form factors \cite{Ahmady:2016ufq} as well as the pion PDF after taking into account perturbative QCD evolution \cite{Ahmady:2018muv}. We shall use our spin-improved pion holographic wavefunction, without any further adjustment of its parameters, in order to predict the twist-$2$ pion TMDs. Recently, in Ref. \cite{Bacchetta:2017vzh}, the unpolarized pion TMD, $f_{1}(x, k_\perp)$, was predicted using the original pion holographic wavefunction (i.e. to which our spin-improved wavefunction reduces when $B=0$). Here, we go beyond the analysis in Ref. \cite{Bacchetta:2017vzh} by also predicting the holographic Boer-Mulders function (which indeed vanishes if $B=0$).

\section{TMDs}

 The pion TMDs are derived from the quark correlation function 
 \begin{equation}
 	\Phi_{ij}^{[\Gamma]} (x,\mathbf{k})= \int \frac{d z^- d^2 z_{\perp}}{2\pi (2\pi)^2} e^{i z\cdot k}  \langle \pi | \bar{\Psi}_j(0) \Gamma \mathcal{L}^{\dagger}(\mathbf{0}|n) \mathcal{L} (\mathbf{z}|n)\Psi_i(z) | \pi \rangle_{z^+=0} 
 \label{Correlator}
 \end{equation}
 where
 \begin{equation}
 	\mathcal{L}_{A^+=0}(\mathbf{z}_\perp|n)= \mathcal{P} \exp\left(-ig \int_{\mathbf{z}_\perp}^{\infty} \mathrm{d} \mathbf{\eta}_\perp \cdot \mathbf{A}_\perp(\eta^-=n \cdot \infty,\mathbf{z}_{\perp})\right)
 \label{gauge-link}
 \end{equation}
 is the gauge link (in the light-front gauge, $A^+=0$) which guarantees colour gauge invariance and $n=(0,+1(-1),0)$ in SIDIS (DY). The unpolarized TMD and Boer-Mulders function are given by  \cite{Pasquini:2014ppa}
  \begin{equation}
 	f_{1}(x, k_\perp)=  \frac{1}{2} \mathrm{Tr}(\Phi ^{[\gamma^+]})  	
 	\label{fTMD-def} 
 \end{equation}
 and 
 \begin{equation}
 	 h^\perp_{1}(x, k_\perp)=  \frac{\epsilon^{ij}\mathbf{k}^jM_\pi}{2k_\perp^2} \mathrm{Tr} (\Phi^{[i \sigma^{i+}\gamma_5]})  	
 	 \label{BM-TMD-def}  
 \end{equation} 
 respectively.

 Ignoring the gauge link,
 \begin{equation}
\mathrm{Tr} (\Phi^{[\Gamma]}) = \sum_{h,\bar{h}, h^\prime} \frac{1}{16\pi^3 k^{+}} \Psi_{h^\prime \bar{h}}^*(x,\mathbf{k}) \Psi_{h\bar{h}}(x,\mathbf{k}) \bar{u}_{h^\prime}(k^{+},\mathbf{k}) \Gamma u_h (k^+,\mathbf{k})	\;.
\label{Tr-LO}
\end{equation} 
 Using the light-front matrix element \cite{Lepage:1980fj}
 \begin{equation}
 	\bar{u}_{h^\prime}(k^+,\mathbf{k}) \gamma^+ u_h(k^+,\mathbf{k})=2 k^{+} \delta_{h h^\prime} \;,
 \end{equation}
  it follows that 
  \begin{equation}
 	f_{1}(x,  k_\perp)=\frac{1}{16\pi^3} \sum_{h,\bar{h}} |\Psi_{h \bar{h}}(x,\mathbf{k})|^2  \;.
 \label{hf1}
 \end{equation} 
 Thus the ordinary PDF
  \begin{equation}
 	f(x)=\int \mathrm{d}^2 \mathbf{k} f_1(x, k_\perp) \;,
 \label{PDF}
 \end{equation}
 satisfies the normalization condition
 \begin{equation}
 	\int \mathrm{d} x f(x)=1  \;.
 \label{PDF-norm}
 \end{equation}
 This embodies the assumption that our holographic distributions are valid at a low hadronic scale where there are only valence quarks and no sea quarks and gluons in the pion. In Ref. \cite{Ahmady:2018muv}, we evolved the holographic PDF perturbatively in order to fit the re-analyzed E615 data \cite{Aicher:2010cb}. 
   
 On the other hand, using the light-front matrix element,
 \begin{equation}
 	\bar{u}_{h^\prime}(k^+,\mathbf{k}) \epsilon^{ij} k^j \sigma^{i+} \gamma^5 u_h(k^+,\mathbf{k})=2 k^{+} h^\prime \delta_{-h^\prime h} k_\perp e^{-ih^\prime \theta_{k_\perp}} 
 \end{equation}
 in Eq. \eqref{Tr-LO}, we deduce that the Boer-Mulders function vanishes. This is because
\begin{equation}
\sum_{h,\bar{h}} \Psi_{-h \bar{h}}^*(x,\mathbf{k})h k_\perp e^{i h \theta_{k_\perp}}  \Psi_{h\bar{h}}(x,\mathbf{k}) =0 \;,
\label{zeroBM}
\end{equation}
as can be readily verified using our spin-improved holographic wavefunctions. To generate a non-zero Boer-Mulders function, we need to take into account the gauge link. Physically, this is equivalent to taking into account initial or final state interactions of the active quark with the target remnant, which we refer collectively as gluon rescattering. We assume that this physics is encoded in a gluon rescattering kernel $G (x, \mathbf{k}- \mathbf{k}^\prime)$ such that 
  \begin{equation}
\mathrm{Tr} (\Phi^{[\Gamma]}) = \sum_{h,\bar{h}, h^\prime} \int \frac{\mathrm{d}^2 \mathbf{k}^\prime}{16\pi^3 k^{\prime +}} G (x, \mathbf{k}- \mathbf{k}^\prime)  
\Psi_{h^\prime \bar{h}}^*(x,\mathbf{k}^\prime) \Psi_{h\bar{h}}(x,\mathbf{k}) \bar{u}_{h^\prime}(k^{\prime +},\mathbf{k}^\prime)  \Gamma u_h (k^+,\mathbf{k})	\;.
\label{Tr}
\end{equation}
Using the fact that
 \begin{equation}
 	\bar{u}_{h^\prime}(k^{\prime +},\mathbf{k}^\prime)i \epsilon^{ij} k^j \sigma^{i+} \gamma^5 u_h (k^+,\mathbf{k})=2 i k^{\prime +} h^\prime \delta_{-h^\prime h} k_\perp e^{-ih^\prime \theta_{k_\perp}} 
 \end{equation}
it follows that
\begin{equation}
	\frac{1}{2}\mathrm{Tr} (\Phi^{[i \sigma^{i+}\gamma_5]}) =\int \frac{\mathrm{d}^2 \mathbf{k}^\prime}{16\pi^3}~i G(x, \mathbf{k}- \mathbf{k}^\prime) \sum_{h,\bar{h}} \Psi_{-h,\bar{h}}^*(x, \mathbf{k}^\prime) h k_\perp e^{i h \theta_{k_\perp}} \Psi_{h,\bar{h}}(x,\mathbf{k}) \;, 
\label{Tr-iG}
\end{equation}
and therefore Eq. \eqref{BM-TMD-def} yields
 \begin{equation}
	k_\perp^2 h_{1}^{\perp}(x, k_\perp^2) = M_\pi \int \frac{\mathrm{d}^2 \mathbf{k}^\prime}{16\pi^3}~ i G(x, \mathbf{k}- \mathbf{k}^\prime) \sum_{h,\bar{h}} \Psi_{-h,\bar{h}}^*(x, \mathbf{k}^\prime) h k_\perp e^{i h \theta_{k_\perp}} \Psi_{h,\bar{h}}(x,\mathbf{k}) \;.
\label{BM-overlap}
\end{equation}
Defining $\mathbf{q}=\mathbf{k}-\mathbf{k}^\prime$, we can rewrite Eq. \eqref{BM-overlap} as
 \begin{equation}
	k_\perp^2 h_{1}^{\perp}(x, k_\perp^2) = M_\pi \int \frac{\mathrm{d}^2 \mathbf{q}}{16\pi^3}~ i G(x, q_\perp) \sum_{h,\bar{h}} \Psi_{-h,\bar{h}}^*(x, \mathbf{k}-\mathbf{q}) h k_\perp e^{i h \theta_{k_\perp}} \Psi_{h,\bar{h}}(x,\mathbf{k}) \;,
\label{BM-overlap-q}
\end{equation}
where we have assumed that $G(x, \mathbf{q})= G(x, q_\perp)$. To proceed, we must specify the form of the gluon rescattering kernel $G(x, q_\perp)$.

 \section{The gluon rescattering kernel}
 
 The simplest approach is to assume that \cite{Brodsky:2002rv,Brodsky:2002cx}
\begin{equation}
 	\Im \mathrm{m} G^{\mathrm{pert.}}(x, q_\perp) \propto \frac{C_F\alpha_s}{q^2_\perp} \;,
 	\label{pert-G}
\end{equation}
referred to as the perturbative Abelian gluon rescattering kernel since it can be derived by working with perturbative Abelian gluons, followed by the replacement $g^2 \to 4\pi C_F \alpha_s$. By hypothesis, the coupling is weak, i.e. $g^2 \ll 1$ which implies that $\alpha_s \ll 0.95$. Yet, there is no consensus in the literature on what value of $\alpha_s$ should be used in Eq. \eqref{pert-G}. For instance, while Ref. \cite{Lu:2004hu} uses $\alpha_s=0.3$,  other authors prefer to use much larger values of $\alpha_s$: $\alpha_s=0.911$  in Ref. \cite{Wang:2017onm}, and $\alpha_s=1.2$ in Ref. \cite{Pasquini:2014ppa}.  Strictly speaking, using such large values of $\alpha_s$ contradicts the weak coupling hypothesis leading to Eq. \eqref{pert-G}. However, taking $\alpha_s \sim 1$ in the perturbative kernel may perhaps be considered as a phenomenological way to account, at least to some extent, for non-perturbative effects.
  
Having said that, it is still clear that Eq. \eqref{pert-G} has the shortcoming of diverging as $q_\perp\to 0$. While this divergence may be regulated when computing the Boer-Mulders function, it remains true that the perturbative kernel might not capture accurately the dynamics of soft gluons which are primarily responsible for generating a non-perturbative quantity like the Boer-Mulders function. In addition, the prescription $g^2 \to 4\pi C_F \alpha_s$ in an Abelian theory  neglects the contribution of crossed gluon ladder diagrams in QCD although the latter are subleading only in a large $N_c$ approximation. An exact non-perturbative computation gluon rescattering kernel is yet not available and, in practice, some approximation scheme is necessary. In Ref. \cite{Gamberg:2009uk}, Gamberg and Schlegel obtained the so-called QCD lensing function \cite{Burkardt:2007xm} from the eikonal amplitude for quark-antiquark scattering via the exchange of both direct and crossed ladder diagrams of non-Abelian soft gluons. In their antiquark spectator framework, the lensing function, $I(x, q_\perp)$, connects the first moment of the Boer-Mulders function with the chiral-odd pion Generalized Parton Distribution (GPD):
\begin{equation}
	M_\pi^2 h_1^{\perp (1)}(x)= \int \frac{\mathrm{d}^2 \mathbf{q}}{2 (2\pi)^2}  q_\perp I(x, q_\perp) \mathcal{H}_1^{\pi}\left(x,-\left(\frac{q_\perp}{\bar{x}}\right)^2\right)
\label{relation-moment-GPD-Gamberg}
\end{equation} 
where the first moment of the Boer-Mulders function is 
\begin{equation}
	2 M_\pi^2 h_1^{\perp (1)}(x)= \int \mathrm{d}^2 \mathbf{k} k_\perp^2 h_{1}^{\perp}(x, k_\perp^2)
	\label{momentBM}
\end{equation}
and the chiral-odd pion GPD is given by
\begin{equation}
	\mathcal{H}_1^{\pi}(x,-\Delta_\perp^2)= \frac{\epsilon^{ij} \Delta^i M_\pi}{2 \Delta_\perp^2}  \int \frac{\mathrm{d}z^-}{2\pi} e^{ik^+z^-} \langle P^+, \mathbf{\Delta} | \bar{\Psi} (0) \sigma^{i+} \gamma_5 \Psi (z) | P^+, \mathbf{0} \rangle_{z^+=0} 
\label{GPD-def}
\end{equation} 
with $\mathbf{\Delta}=-\mathbf{q}/\bar{x}$. As noted in Ref. \cite{Gamberg:2009uk} and proved in Ref. \cite{Meissner:2008ay}, Eq. \eqref{relation-moment-GPD-Gamberg} is not model-independent. Thus, to be able to use the lensing function of Ref. \cite{Gamberg:2009uk}, we must first demonstrate that a factorization of the type given by Eq. \eqref{relation-moment-GPD-Gamberg} also holds in the overlap representation with our spin-improved holographic light-front wavefunctions. 

Inserting Eq. \eqref{BM-overlap} in Eq. \eqref{momentBM} and changing variable  $\mathbf{k} \to \mathbf{q}=\mathbf{k} - \mathbf{k}^\prime$, we obtain
\begin{equation}
	M_\pi^2 h_1^{\perp (1)}(x)=\frac{M_\pi}{2} \int \mathrm{d}^2 \mathbf{q} ~ i G(x, q_\perp)  \int \frac{\mathrm{d}^2 \mathbf{k}^\prime}{16\pi^3} \sum_{h,\bar{h}} \Psi_{-h,\bar{h}}^*(x, \mathbf{k}^\prime) h (q_\perp e^{i h \theta_{q_\perp}} +  k^\prime_\perp e^{i h \theta_{k^\prime_\perp}})\Psi_{h,\bar{h}}(x,\mathbf{k}^\prime+\mathbf{q}),
	\label{moment}
	\end{equation}
which can be re-expressed as
\begin{equation}
	M_\pi^2 h_1^{\perp (1)}(x)=\frac{M_\pi}{2} \int \mathrm{d}^2 \mathbf{q} ~ i  G(x, q_\perp) [q_\perp + F(x, q_\perp)] \int \frac{\mathrm{d}^2 \mathbf{k}^\prime}{16\pi^3} \sum_{h,\bar{h}} \Psi_{-h,\bar{h}}^*(x, \mathbf{k}^\prime) h  e^{i h \theta_{q_\perp}}\Psi_{h,\bar{h}}(x,\mathbf{k}^\prime+\mathbf{q}) \;,
	\label{Re-expressed-moment}
		\end{equation}
		where the function
\begin{equation}
	F(x, q_\perp; \alpha)= \frac{\int \mathrm{d}^2 \mathbf{k}^\prime\sum_{h,\bar{h}} \Psi_{-h,\bar{h}}^*(x, \mathbf{k}^\prime) h k^\prime_\perp e^{i h \theta_{k^\prime_\perp}}\Psi_{h,\bar{h}}(x,\mathbf{k}^\prime+\mathbf{q})}{\int \mathrm{d}^2 \mathbf{k}^\prime \sum_{h,\bar{h}} \Psi_{-h,\bar{h}}^*(x, \mathbf{k}^\prime) h e^{i h \theta_{q_\perp}}\Psi_{h,\bar{h}}(x,\mathbf{k}^\prime + \mathbf{q})} 
\label{second-term}
\end{equation}
depends, \textit{\`a priori}, on $x$ and $q_\perp$ as well as $\alpha$, the set of parameters appearing in the wavefunctions (here $\alpha=\{\kappa,m_f\}$).  However, it turns out that an explicit evaluation of Eq. \eqref{second-term} using our spin-improved holographic wavefunctions, yields
\begin{equation}
	F(q_\perp)=-\frac{q_\perp}{2} \;.
\label{Fq}
\end{equation}
Hence, we can write
\begin{equation}
	M_\pi^2 h_1^{\perp (1)}(x)= \int \frac{\mathrm{d}^2 \mathbf{q}}{2}   \frac{i G(x, q_\perp)}{2} q_\perp \left[M_\pi \int \frac{\mathrm{d}^2 \mathbf{k}^\prime}{16\pi^3} \sum_{h,\bar{h}} \Psi_{-h,\bar{h}}^*(x, \mathbf{k}^\prime) h  e^{i h \theta_{q_\perp}}\Psi_{h,\bar{h}}(x,\mathbf{k}^\prime+\mathbf{q}) \right] \;,
\label{square}
\end{equation}
where the quantity in the square brackets is essentially the overlap representation of the chiral-odd pion GPD. Indeed, using Eq. \eqref{GPD-def}, we are able to show that
\begin{equation}
\frac{q_\perp}{\bar{x}} \mathcal{H}_1^{\pi}\left(x,-\left(\frac{q_\perp}{\bar{x}}\right)^2\right)=-M_\pi \int \frac{\mathrm{d}^2 \mathbf{k}^\prime}{16\pi^3} \sum_{h,\bar{h}} \Psi_{-h,\bar{h}}^*(x, \mathbf{k}^\prime) h  e^{i h \theta_{q_\perp}}\Psi_{h,\bar{h}}(x,\mathbf{k}^\prime+\mathbf{q}) \;.
\end{equation}
Thus, we can rewrite Eq. \eqref{square} as
\begin{equation}
M_\pi^2 h_1^{\perp (1)}(x)= -\int \frac{\mathrm{d}^2 \mathbf{q}}{2}  \frac{i G(x, q_\perp)}{2} \frac{q^2_\perp}{\bar{x}} \mathcal{H}_1^{\pi}\left(x,-\left(\frac{q_\perp}{\bar{x}}\right)^2\right) \;.
\label{relation-moment-GPD}
\end{equation}
We have thus shown that the first moment of our holographic Boer-Mulders function can indeed be expressed as a convolution of the chiral odd pion GPD with the gluon rescattering kernel: a factorization analogous to Eq. \eqref{relation-moment-GPD-Gamberg}. This allows us to compare Eqs. \eqref{relation-moment-GPD} and \eqref{relation-moment-GPD-Gamberg} and deduce that
\begin{equation}
i G(x, q_\perp)= -\frac{2}{(2\pi)^2} \frac{\bar{x} I(x, q_\perp)}{q_\perp} \;.
\label{Relation-GI}
\end{equation}

\section{The lensing function}

In Ref. \cite{Gamberg:2009uk}, the lensing function is derived for final state rescattering by soft U(1), SU(2) and SU(3) gluons. In all three cases, the lensing function is negative and its magnitude increases with $N_c$. In impact space, the lensing function is given by
\begin{equation}
	\mathcal{I}(x, b_\perp)=\frac{\bar{x}}{2N_c} \frac{\chi^\prime}{4} C \left(\frac{\chi}{4}\right)
	\label{b-space-lensing-function}
\end{equation}
where 
\begin{equation}
	C\left(\frac{\chi}{4}\right)=\mathrm{Tr} \left\{ \Im \mathrm{m} \mathrm{f}^\prime \left(\frac{\chi}{4}\right) + \frac{1}{2}\left[ \Im \mathrm{m} f^\prime \left(\frac{\chi}{4}\right) \Re \mathrm{e} f\left(\frac{\chi}{4}\right) -\Im \mathrm{m} f \left(\frac{\chi}{4}\right) \Re \mathrm{e} f^\prime\left(\frac{\chi}{4}\right)\right]\right\}
	\label{color-function}
	\end{equation}
is a colour function and
 \begin{equation}
  	\chi \left(\frac{b_\perp}{\bar{x}}\right)=\frac{g^2}{2\pi} \int \mathrm{d} k_\perp k_\perp J_0 \left(\frac{b_\perp}{\bar{x}} k_\perp \right) \mathcal{D}_1(-k_\perp^2) 
  \label{eikonal-phase}
  \end{equation}
is the eikonal phase with $\mathcal{D}_1(-k_\perp^2)$ being the gauge-independent part of the gluon propagator. The momentum space lensing function is given by the inverse Fourier transform of Eq. 17 in Ref. \cite{Gamberg:2009uk}, i.e.
\begin{equation}
 	I(x, q_\perp) \frac{\mathbf{q}^i}{q_\perp}=-\frac{i}{\bar{x}^3} \int \mathrm{d}^2 \mathbf{b} \exp\left(-i\frac{\mathbf{q} \cdot \mathbf{b}}{\bar{x}}\right) \mathcal{I}(x, b_\perp) \frac{\mathbf{b}^i}{b_\perp} \;.
 \label{FT-I}
 \end{equation}
The real and imaginary parts of $f(\chi/4)$ in Eq. \eqref{color-function} originate from the real and imaginary parts of the eikonal amplitude for quark-antiquark scattering via the exchange of generalized infinite ladders of gluons. As can be seen, it is the imaginary part of the eikonal amplitude that is responsible for a non-vanishing lensing function. There is also a contribution from the real part of the eikonal amplitude, although, as we shall see, it is subleading in the perturbative limit. 

For U(1) gluons,
\begin{equation}
	\Re \mathrm{e} f^{\mathrm{U(1)}}= \cos \chi -1 
\label{fU1-Re}
\end{equation}
and
\begin{equation}
	\Im \mathrm{m} f^{\mathrm{U(1)}}= \sin \chi  
\label{fU1-Im}
\end{equation}
while for SU(3) gluons,
\begin{equation}
	\Re \mathrm{e} [f^{\mathrm{SU(3)}}_{\alpha \beta} ](a)= \delta_{\alpha \beta} (-c_2 a^2 + c_4 a^4 -c_6a^6-c_8 a^8 + ...)
\label{fSU3-Re}
\end{equation}
and
\begin{equation}
	\Im \mathrm{m} [f^\mathrm{SU(3)}_{\alpha \beta} ](a)= \delta_{\alpha \beta} (c_1 a - c_3 a^3 + c_5a^5-c_7 a^7 + ...)
\label{fSU3-Im}
\end{equation}
where $a \equiv\chi/4$ and $c_i$ are numerical coefficients given in Ref. \cite{Gamberg:2009uk}. 
Eq. \eqref{fU1-Re} to  Eq. \eqref{fSU3-Im} reveal that the real part of $f(\chi/4)$ is subleading for perturbative gluons (since $g^2 \ll 1$ and $\chi \ll 1$).

We can now find the lensing function for perturbative Abelian gluons. To leading $g^2$, the colour function becomes
\begin{equation}
	C^{\mathrm{pert.}}_{\mathrm{U(1)}}(\chi)= 4 \cos \chi \;.
	\label{CU1-pert}
\end{equation}
Now, using the perturbative Feynman gluon propagator,  
\begin{equation}
	\mathcal{D}_1(-k_\perp^2)=\frac{1}{k_\perp^2}
	\label{Feynmann-propagator} 
\end{equation}
in Eq. \eqref{eikonal-phase}, we find that
\begin{equation}
	\chi^\prime \left(\frac{b_\perp}{\bar{x}}\right)=-\frac{g^2}{2\pi} \frac{\bar{x}}{b_\perp} \;.
	\label{derivative-phase}
\end{equation}
Using Eqs. \eqref{derivative-phase} and \eqref{CU1-pert} in Eq. \eqref{b-space-lensing-function}, we find that
\begin{equation}
	\mathcal{I}^{\mathrm{pert.}}_{U(1)}(x, b_\perp)= - \frac{g^2}{4 \pi} \frac{\bar{x}^2}{b_\perp} 
\label{pIU1}
\end{equation}
so that Eq. \eqref{FT-I} yields
\begin{equation}
 	I^{\mathrm{pert.}}_{U(1)}(x, q_\perp)=- \frac{g^2}{2} \frac{\bar{x}} {q_\perp} \;.
\label{pert-I-U1}
\end{equation}
Eq. \eqref{Relation-GI} then tells us that
\begin{equation}
	i G^{\mathrm{pert.}}_{U(1)}(x, q_\perp)= \frac{g^2}{4\pi^2} \frac{1} {q^2_\perp} \;.
\end{equation}
After the replacement $g^2 \to 4\pi C_F \alpha_s$, we obtain
\begin{equation}
	i G^{\mathrm{pert.}}(q_\perp)=  \frac{\alpha_s C_F}{\pi q_\perp^2} 
	\label{pert-G-us}
\end{equation}
which is consistent with Eq. \eqref{pert-G}. Eq. \eqref{pert-G-us} is our perturbative kernel. 
   
For non-perturbative SU(3) gluons, both the real and imaginary parts of $f^{\mathrm{SU(3)}}_{\alpha \beta}(\chi/4)$ contribute to the colour function. To compute the eikonal phase $\chi$, we follow Ref. \cite{Gamberg:2009uk} in using a non-perturbative Dyson-Schwinger  gluon propagator given by 
\begin{equation}
	\mathcal{D}_1 (k_\perp^2, \Lambda_{\mathrm{QCD}}^2)=\frac{1}{k_\perp^2}\left(\frac{\alpha_s(k_\perp^2)}{\alpha_s(\Lambda_{\mathrm{QCD}}^2)}\right)^{1+2\delta}\left( \frac{c(k^2_\perp/\Lambda^2)^\kappa+d(k^2_\perp/\Lambda^2)^{2\kappa}}{1+c(k^2_\perp/\Lambda^2)^\kappa+d(k^2_\perp/\Lambda^2)^{2\kappa}}\right)^2 
\label{DS-gluon}
\end{equation}
with
\begin{equation}
	\alpha_s(\mu^2)=\frac{\alpha_s(0)}{\ln [e + a_1(\mu^2/\Lambda^2)^{a_2} + b_1(\mu^2/\Lambda^2)^{b_2}]} 
\end{equation}
where all parameters are taken from Ref. \cite{Fischer:2003rp} and are explicitly given in Ref. \cite{Gamberg:2009uk}.

\begin{figure}[hbt]
    \includegraphics[width=12cm]{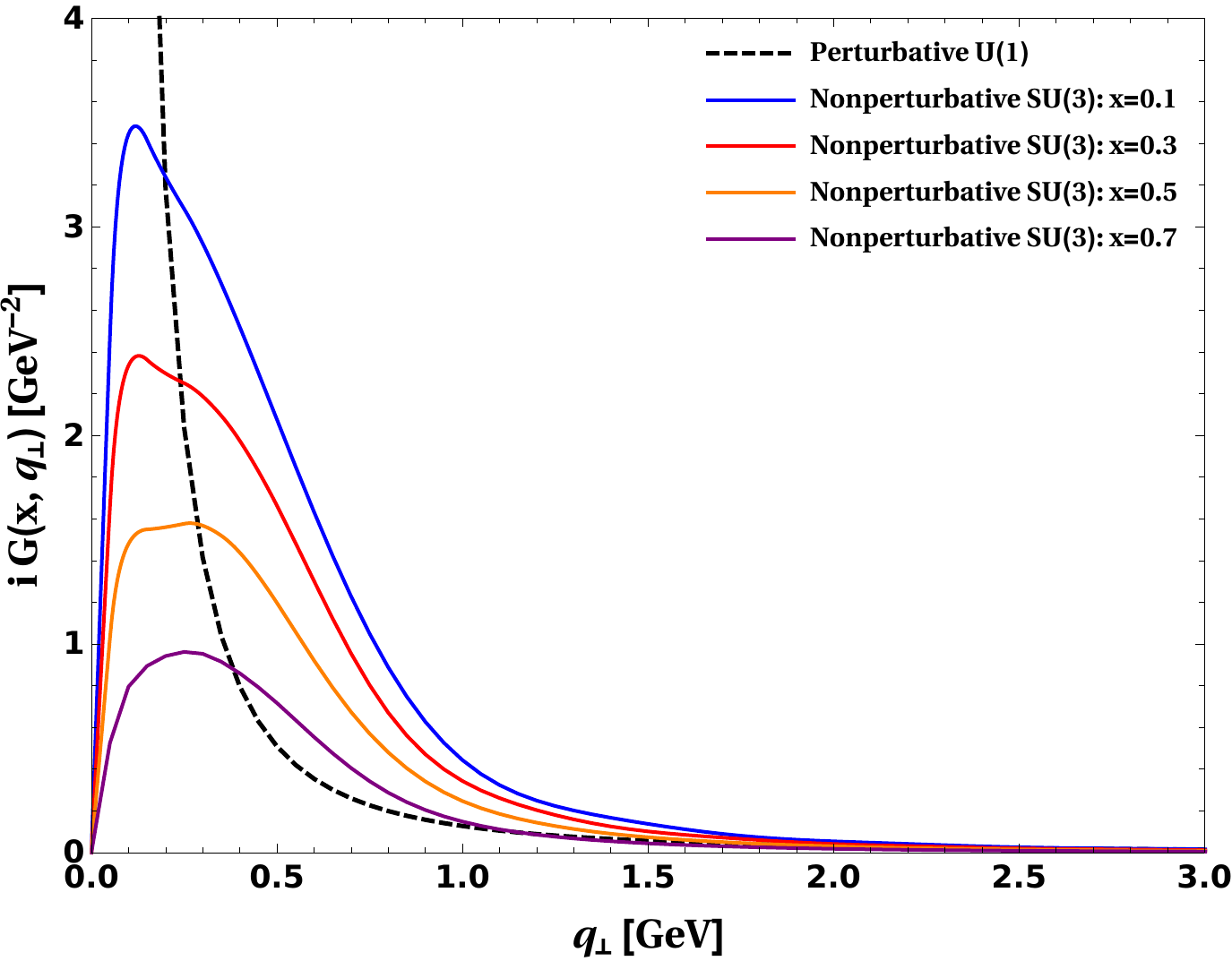}
  \caption{Solid curves: the non-perturbative SU(3) kernel at different values of $x$: $x=0.1$ (blue), $x=0.3$ (red), $x=0.5$ (orange) and $x=0.7$ (purple). Dashed black curve: the perturbative kernel with $\alpha_s=0.3$.}
  \label{fig:G}
  \end{figure}

In Fig. \ref{fig:G}, we compare the perturbative Abelian kernel to the non-perturbative SU(3) kernel. As can seen, the two kernels are very different in the $q_\perp \le 1$ GeV region where the non-perturbative kernel offers the advantage of being infrared finite while peaking at low $q_\perp$. A notable feature of the non-perturbative kernel is that it is not symmetric under $\bar{x} \leftrightarrow x$: its maximum decreases with increasing $x$. On the other hand, the perturbative kernel has no $x$ dependence and, as we noted before, diverges as $q_\perp \to 0$. 

 \section{Holographic TMDs}
 Having specified both the light-front wavefunctions and the gluon rescattering kernel, we are now in a position to find explicit expressions for the holographic pion TMDs. We start with the Boer-Mulders function.  The overlap appearing in Eq. \eqref{BM-overlap-q} is given by
  \begin{equation}
 	\sum_{h,\bar{h}} \Psi_{h,\bar{h}}^*(x,\mathbf{k} - \mathbf{q}) h e^{ih \theta_k}\Psi_{h,\bar{h}}(x,\mathbf{k})= 2 B (M_\pi x\bar{x} + B m_f)  q_\perp \cos (\theta_{q_\perp}-\theta_{k_\perp})\frac{\Psi(x,(\mathbf{k}-\mathbf{q} )^2) \Psi(x, \mathbf{k}^2)}{(x\bar{x})^2} 
 \label{overlap}
 \end{equation}
 where
 \begin{equation}
 \Psi(x,(\mathbf{k}-\mathbf{q} )^2) \Psi(x, \mathbf{k}^2)=\frac{\mathcal{N}^2}{x\bar{x}} \exp \left(-\frac{k_\perp^2 + m_f^2}{\kappa^2 x\bar{x}}\right) \exp\left(-\frac{q_\perp^2}{2\kappa^2 x\bar{x}}\right) \exp\left(\frac{q_\perp k_\perp \cos (\theta_{q_\perp}-\theta_{k_\perp})}{\kappa^2 x\bar{x}}\right)  \;.	 
 \end{equation}
 Inserting Eq. \eqref{overlap} in Eq. \eqref{BM-overlap-q}, and integrating over $\theta_{q_\perp}$, we find that
  \begin{eqnarray}
 	 h_1^\perp (x, k_\perp)&=& B  \frac{M_\pi x\bar{x} + B m_f}{(x\bar{x})^3} \mathcal{N}^2 \frac{M_\pi}{k_\perp}\exp \left(-\frac{k_\perp^2 + m_f^2}{\kappa^2 x\bar{x}}\right) \nonumber  \\ && \times \int \frac{\mathrm{d} q_\perp}{4\pi^2} q_\perp^2 
 	 i G(x, q_\perp) \exp\left(-\frac{q_\perp^2}{2\kappa^2 x\bar{x}}\right) I_1\left(-\frac{k_\perp q_\perp}{\kappa^2 x \bar{x}} \right) 
 	\label{hBM}
 \end{eqnarray}
 where $I_1$ is the modified Bessel function of the first kind. If we use the perturbative gluon rescattering kernel in Eq. \eqref{hBM}, we can obtain an analytic form for the holographic Boer-Mulders function:
  \begin{eqnarray}
  	h_1^{\perp \mathrm{pert.}} (x, k_\perp^2)&=& \alpha_s B C_F \frac{M_\pi \mathcal{N}^2}{4 \pi^3} \frac{M_\pi x\bar{x} + B m_f}{(x\bar{x})^2} \left(\frac{\kappa}{k_\perp} \right)^2 \nonumber \\
  &&	\times \exp \left(-\frac{k_\perp^2 + 2m_f^2}{2\kappa^2 x\bar{x}}\right)\left(1-\exp \left(-\frac{k_\perp^2}{2\kappa^2 x\bar{x}}\right)\right) \;.
  	\end{eqnarray}
 As expected, if $B \to 0$, the holographic Boer-Mulders function vanishes. On the other hand, for $B \ge 1$, it is hardly sensitive to the value of $B$ since the wavefunction normalization constant $\mathcal{N} \sim 1/B^2$ for $B \ge 1$.

\begin{figure}[hbt]
  \includegraphics[width=8cm]{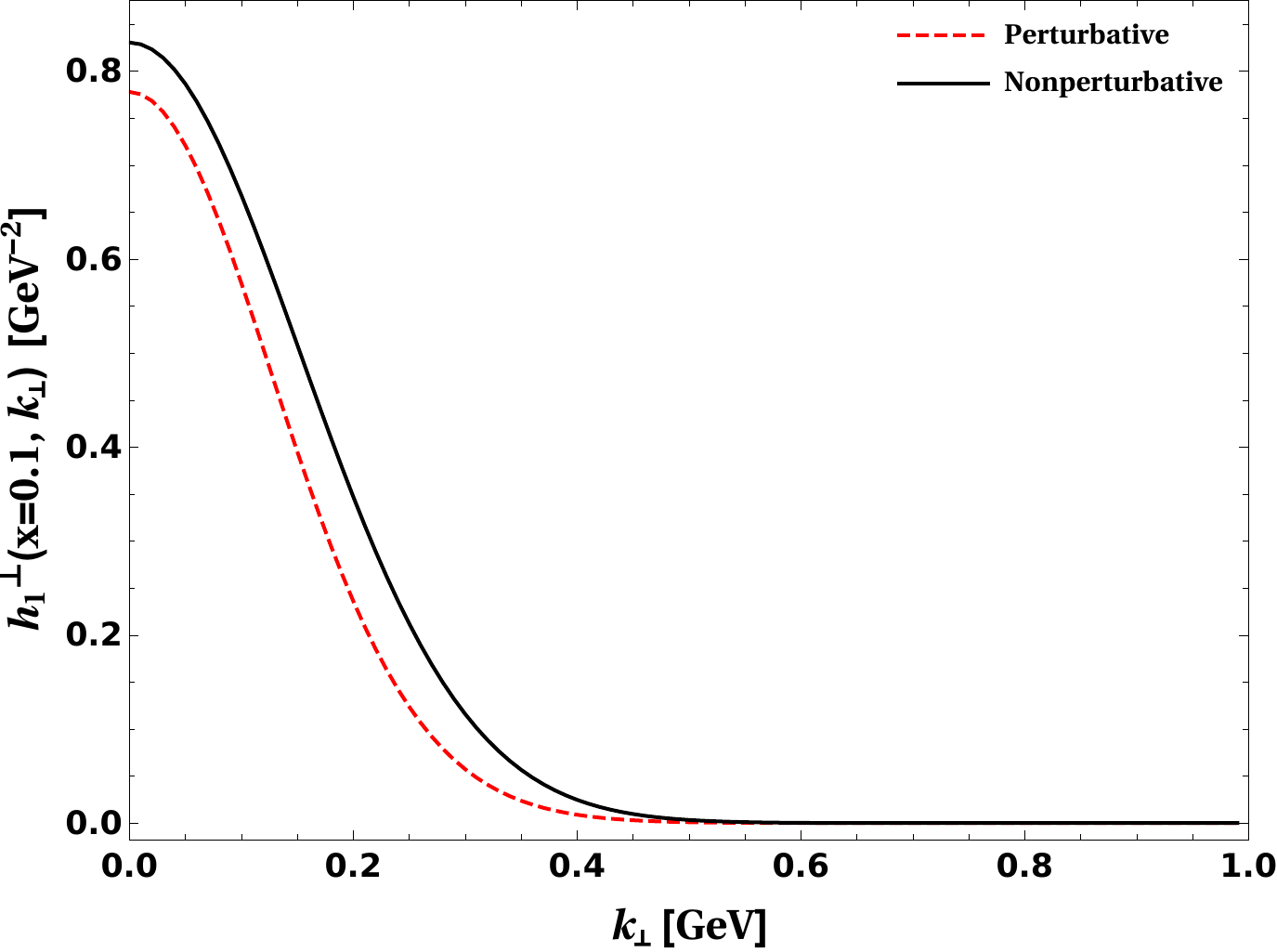}\includegraphics[width=8cm]{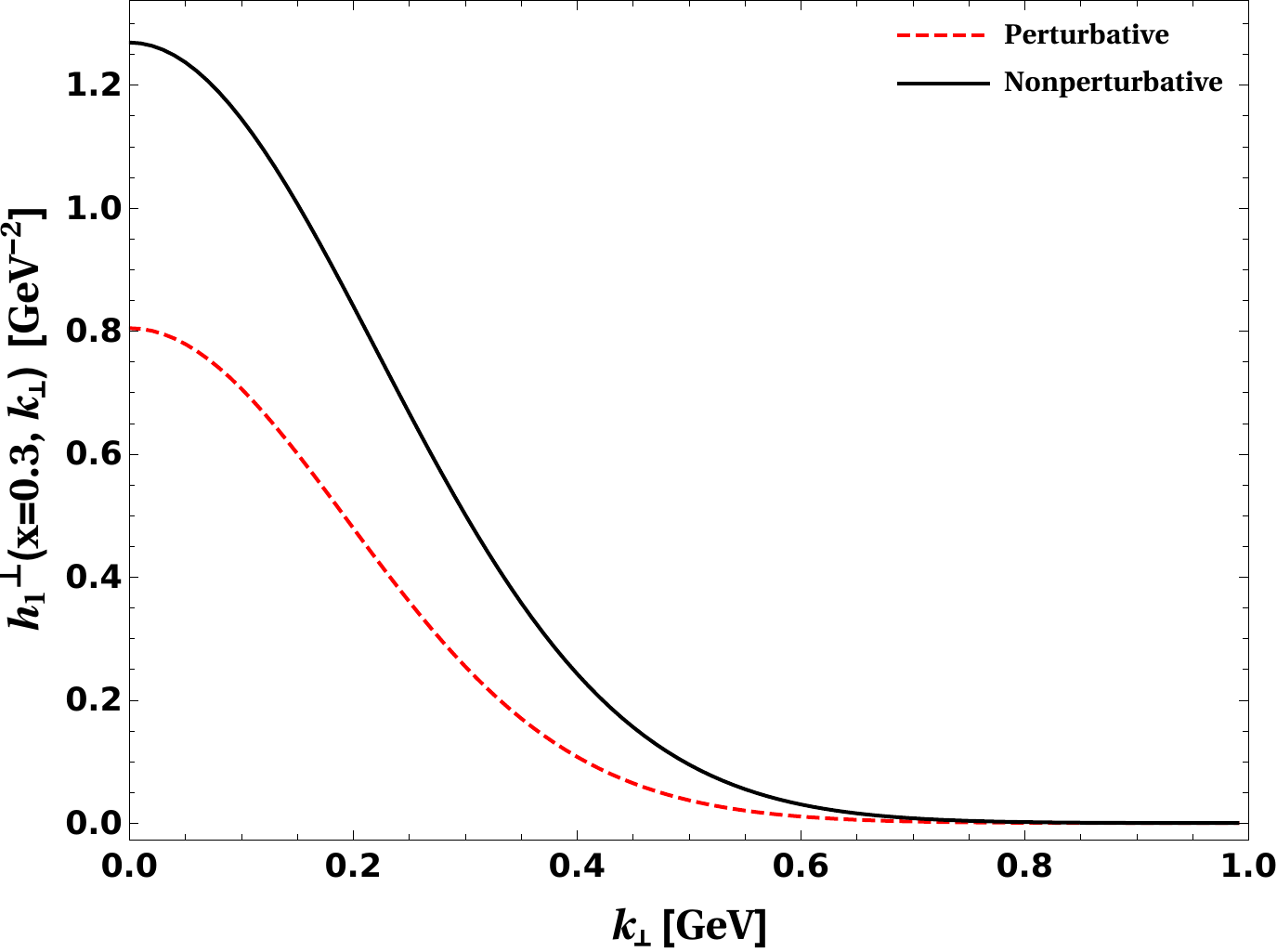}\\
 \includegraphics[width=8cm]{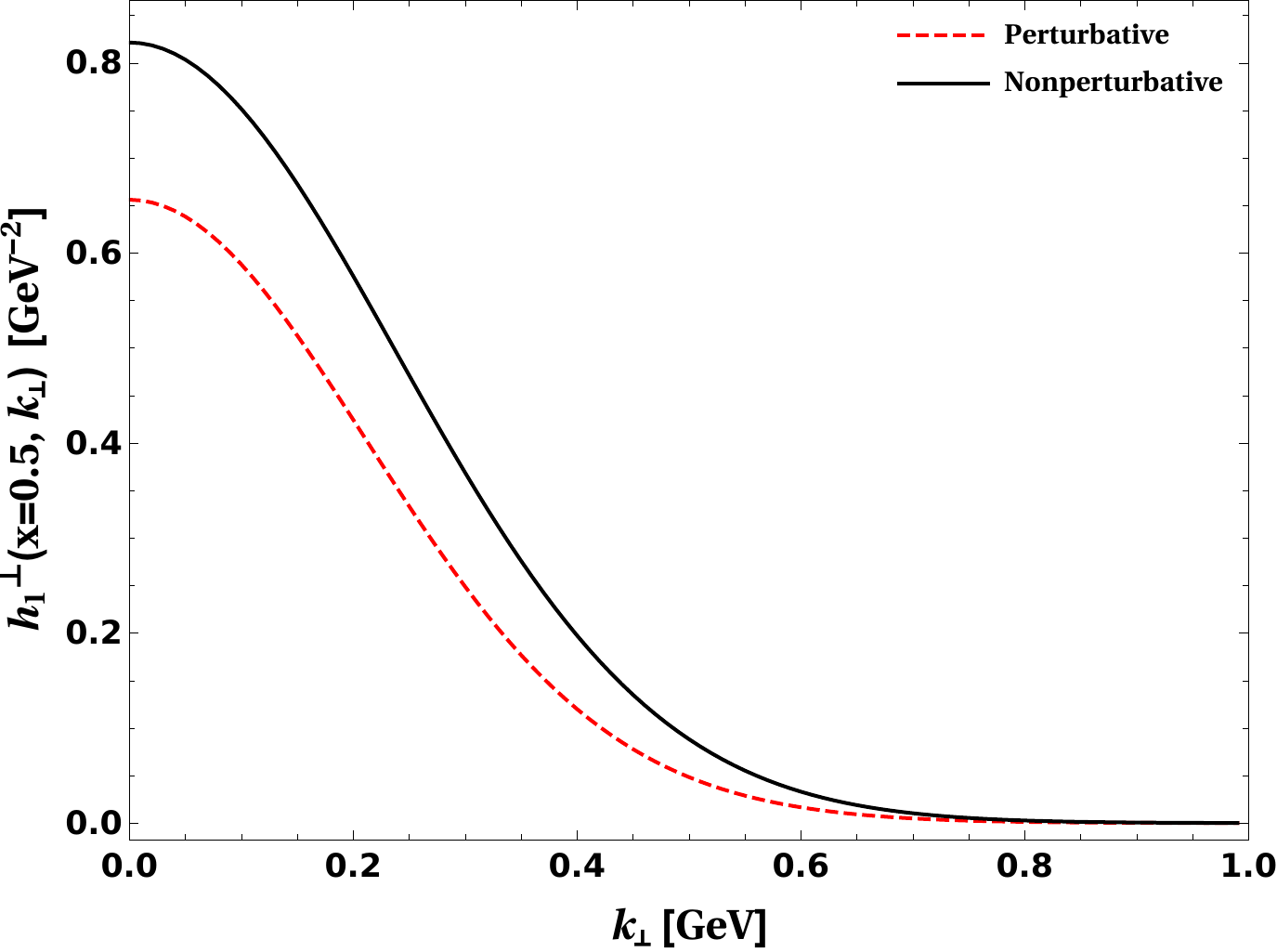} \includegraphics[width=8cm]{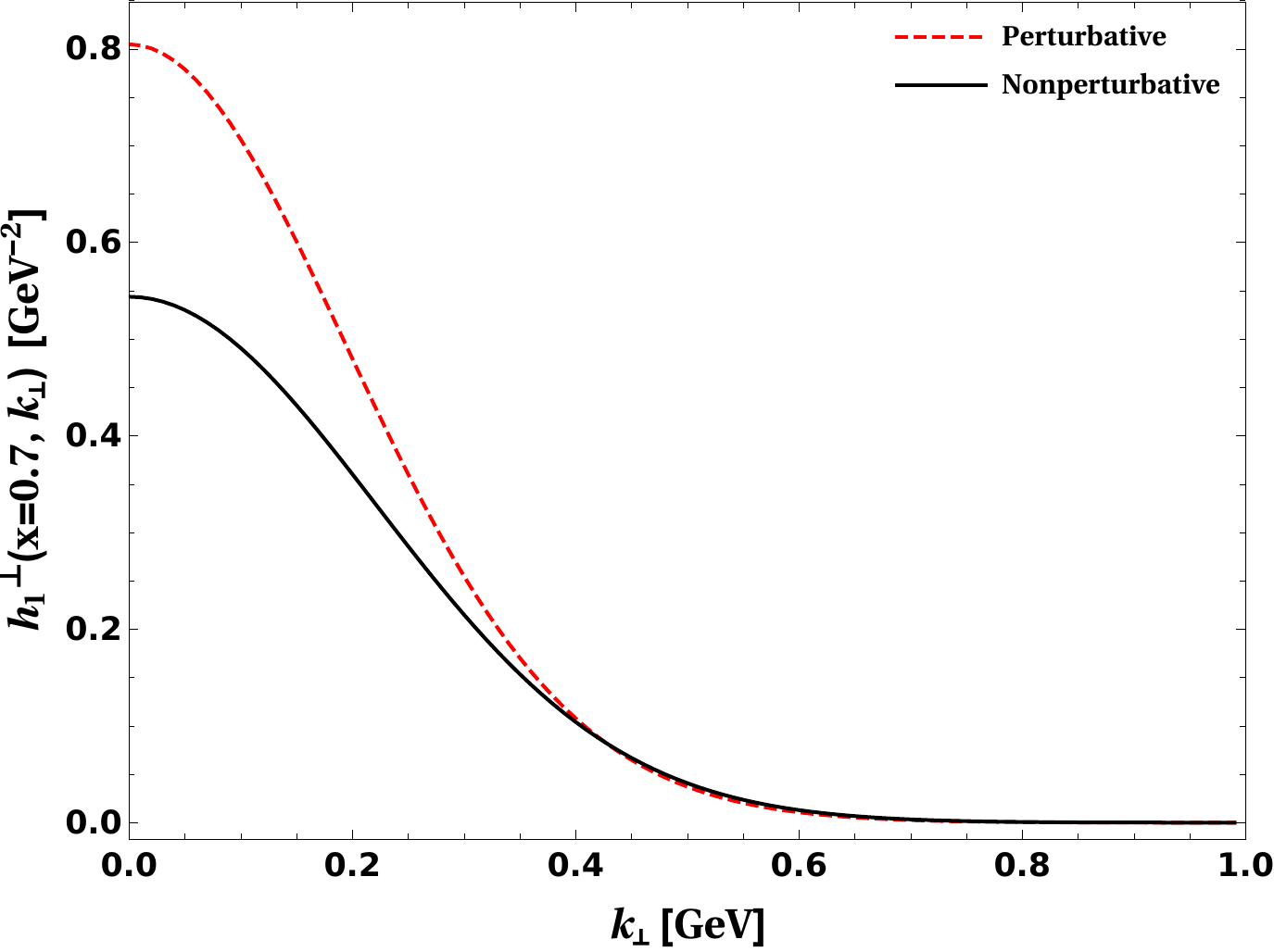}  
 \caption{Solid black curves: the holographic Boer-Mulders function generated by the non-perturbative kernel (solid curves) at different values of $x$: $x=0.1$ (upper left), $x=0.3$ (upper right), $x=0.5$ (lower left) and $x=0.7$ (lower right). Dashed red curves: the holographic Boer-Mulders function generated by the perturbative kernel with $\alpha_s=0.3$.}
  \label{Fig:BM}
  \end{figure}
  
  In Fig. \ref{Fig:BM}, we illustrate the differences between the holographic Boer-Mulders function generated by the perturbative and non-perturbative kernels. As can be seen, a simple rescaling of the normalization of the perturbative kernel, say by increasing $\alpha_s$, cannot fully capture the non-perturbative effects. This is because the difference between the two holographic Boer-Mulders functions is $x$-dependent: at low $x$, the size of the non-perturbatively generated function is larger than that of the perturbatively generated one while the opposite is true at large $x$.

 Let us now give an explicit form for our holographic unpolarized TMD. Using our spin-improved holographic wavefunctions in Eq. \eqref{hf1}, we find that
 \begin{equation}
 	f(x, k_\perp)=\frac{2}{16 \pi^3} \frac{(M_\pi x \bar{x} + B m_f)^2 + B^2 k_\perp^2}{(x\bar{x})^3} \mathcal{N}^2 \exp\left( -\frac{k_\perp^2 + m_f^2}{x\bar{x} \kappa^2}\right) \;.
 \end{equation}
 Contrary to the holographic Boer-Mulders function, our holographic unpolarized TMD does not vanish as $B \to 0$. Instead, it reduces to the original holographic TMD derived in Ref. \cite{Bacchetta:1999kz} with a purely Gaussian dependence on transverse momentum.

\begin{figure}[hbt]
 \includegraphics[width=8cm]{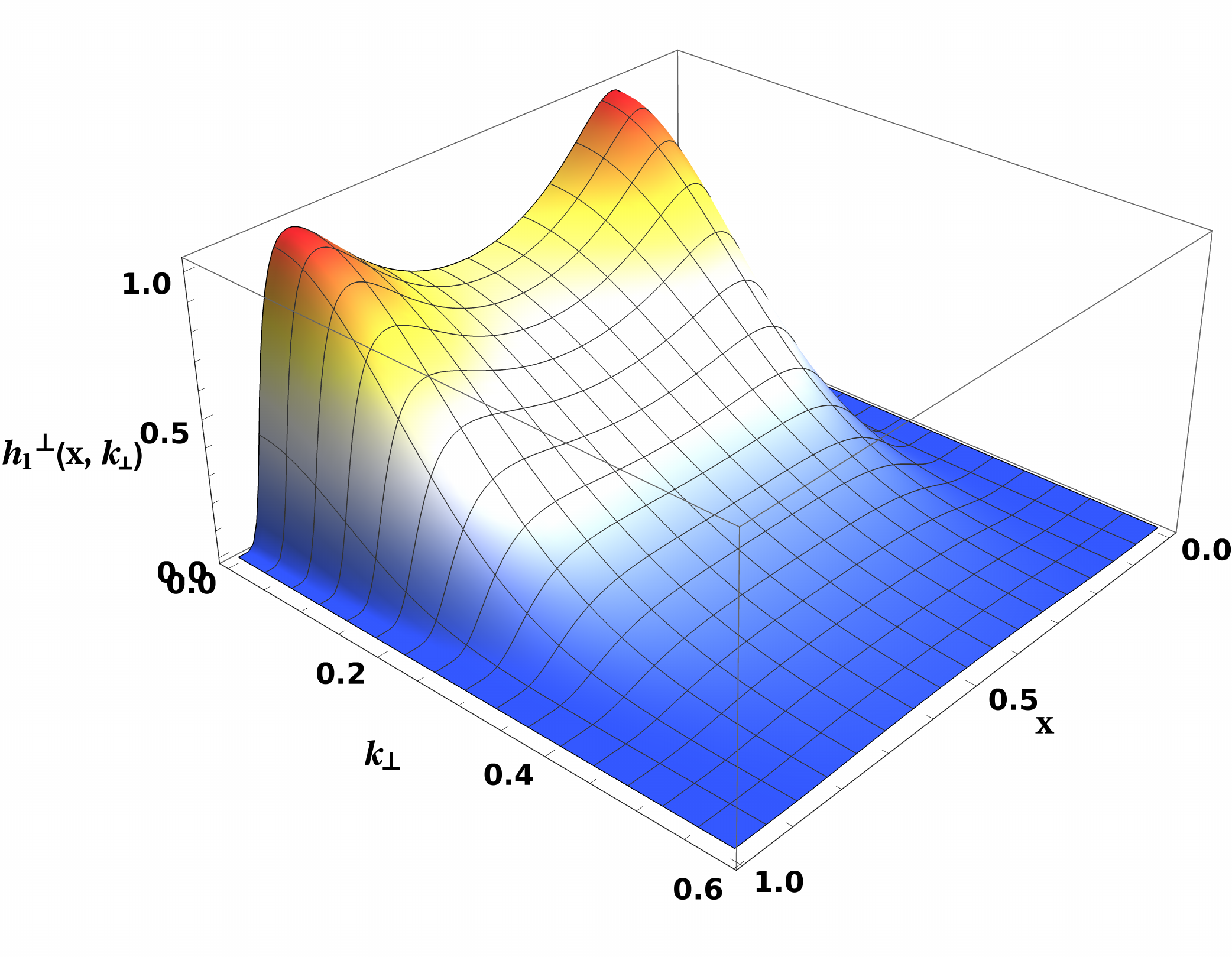}\includegraphics[width=8cm]{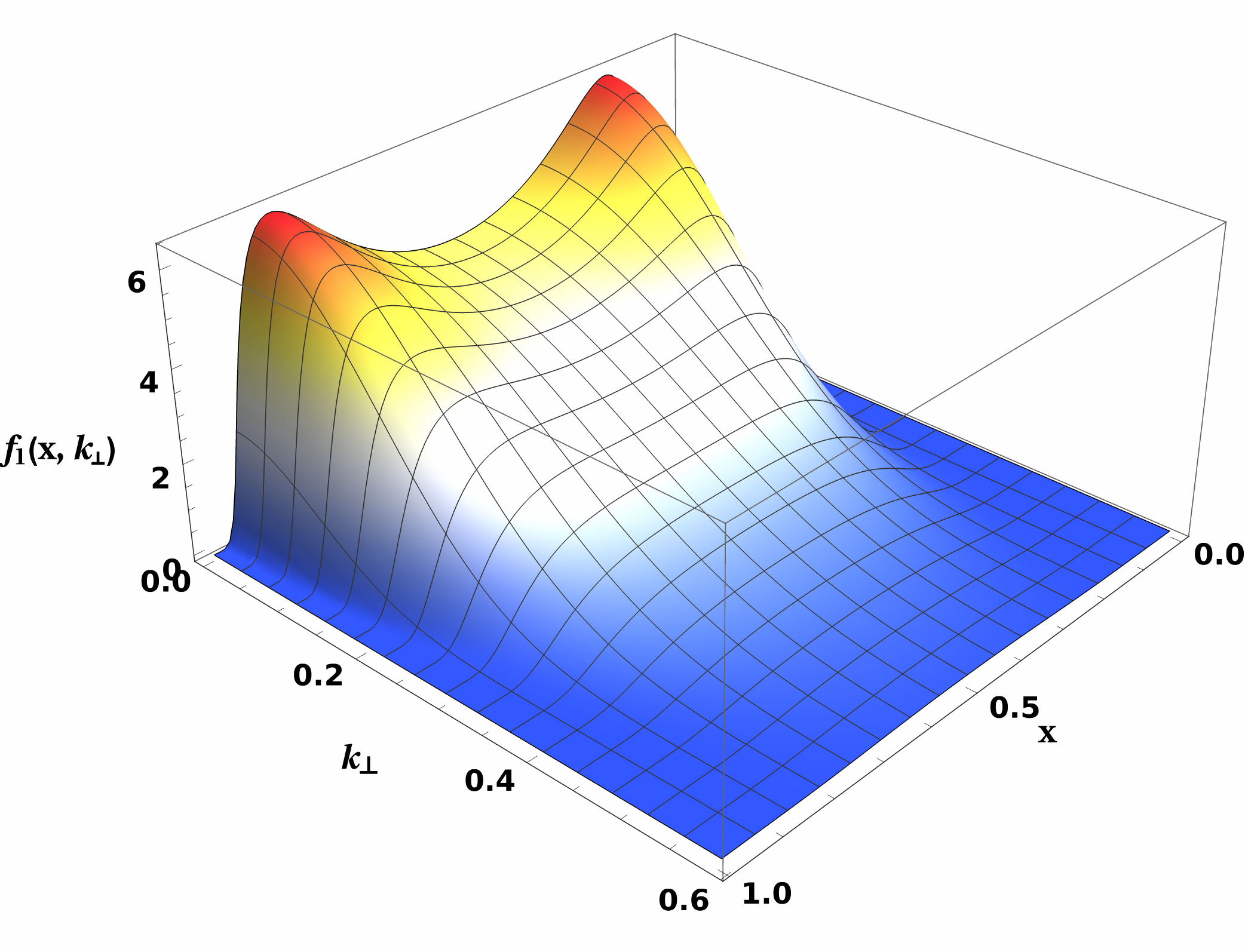}
  \caption{Left: the holographic Boer-Mulders function generated by the perturbative kernel with $\alpha_s=0.3$. Right: the holographic unpolarized TMD. The numbers on the vertical axis are in units of $\mathrm{GeV}^{-2}$ and $k_\perp$ is in GeV.}
   \label{Fig:3d}  
   \end{figure}

 In Fig. \ref{Fig:3d}, we show the 3-dimensional plots of the holographic unpolarized TMD and the holographic Boer-Mulders function generated with the perturbative kernel. The plots reveal a double-humped structure about $x=0.5$ for both holographic TMDs. This feature is inherited from the $x$-dependence of our spin-improved holographic wavefunctions. We note that it does not survive in the holographic Boer-Mulders generated by the non-perturbative kernel since, as we mentioned before, the latter is not symmetric about $x=0.5$. 
   	
\begin{figure}[hbt]
\includegraphics[width=8cm]{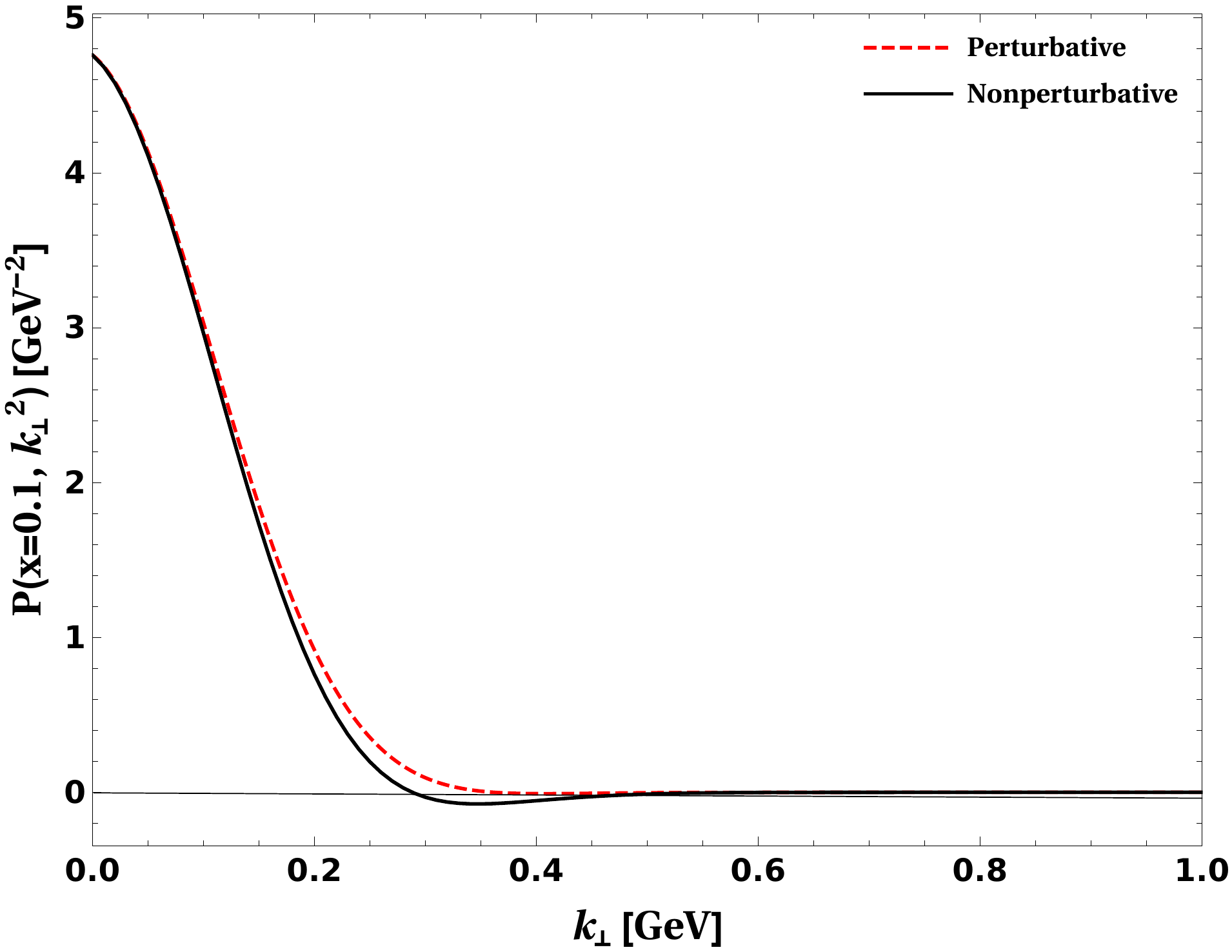}  \includegraphics[width=8cm]{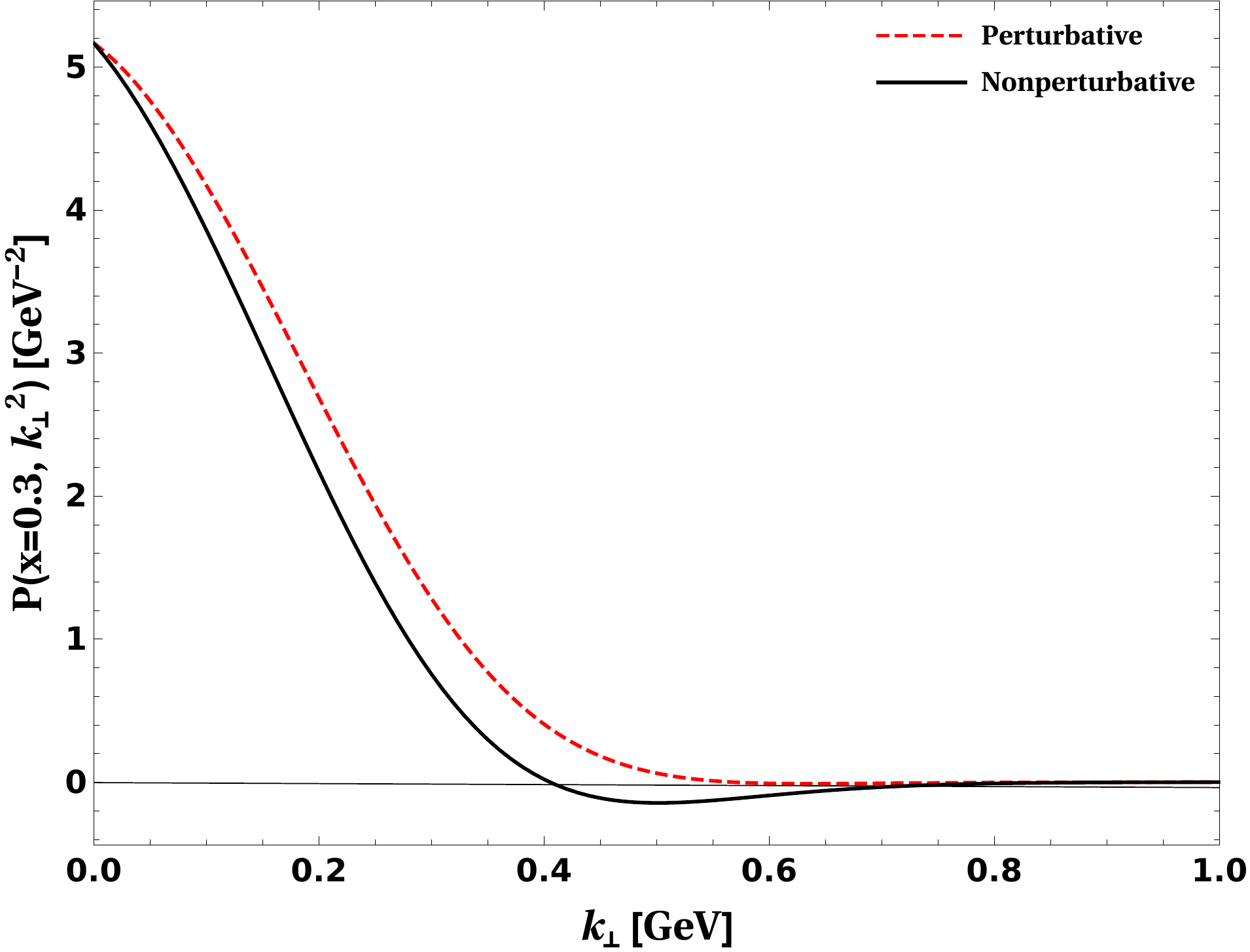}  \\
\includegraphics[width=8cm]{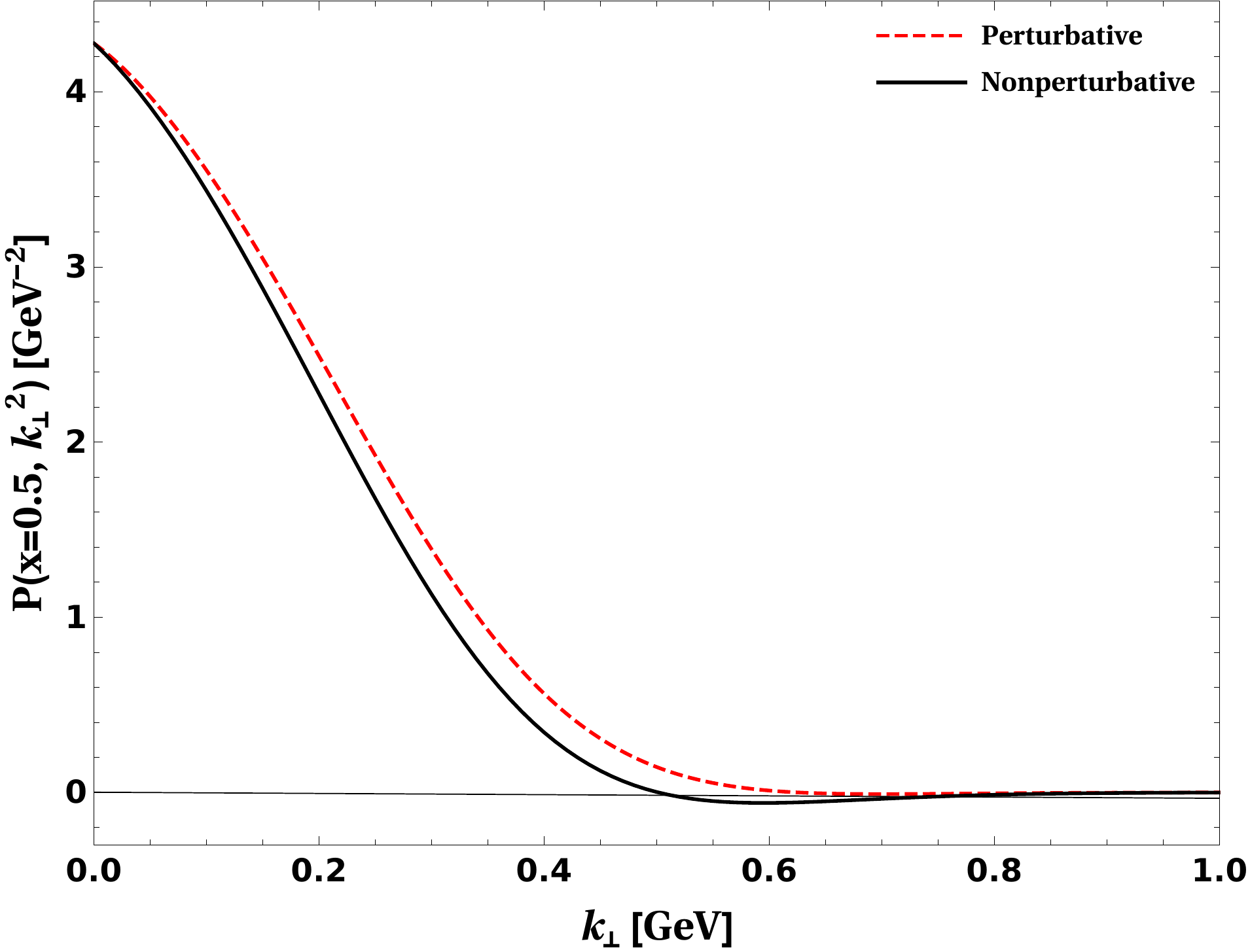} \includegraphics[width=8cm]{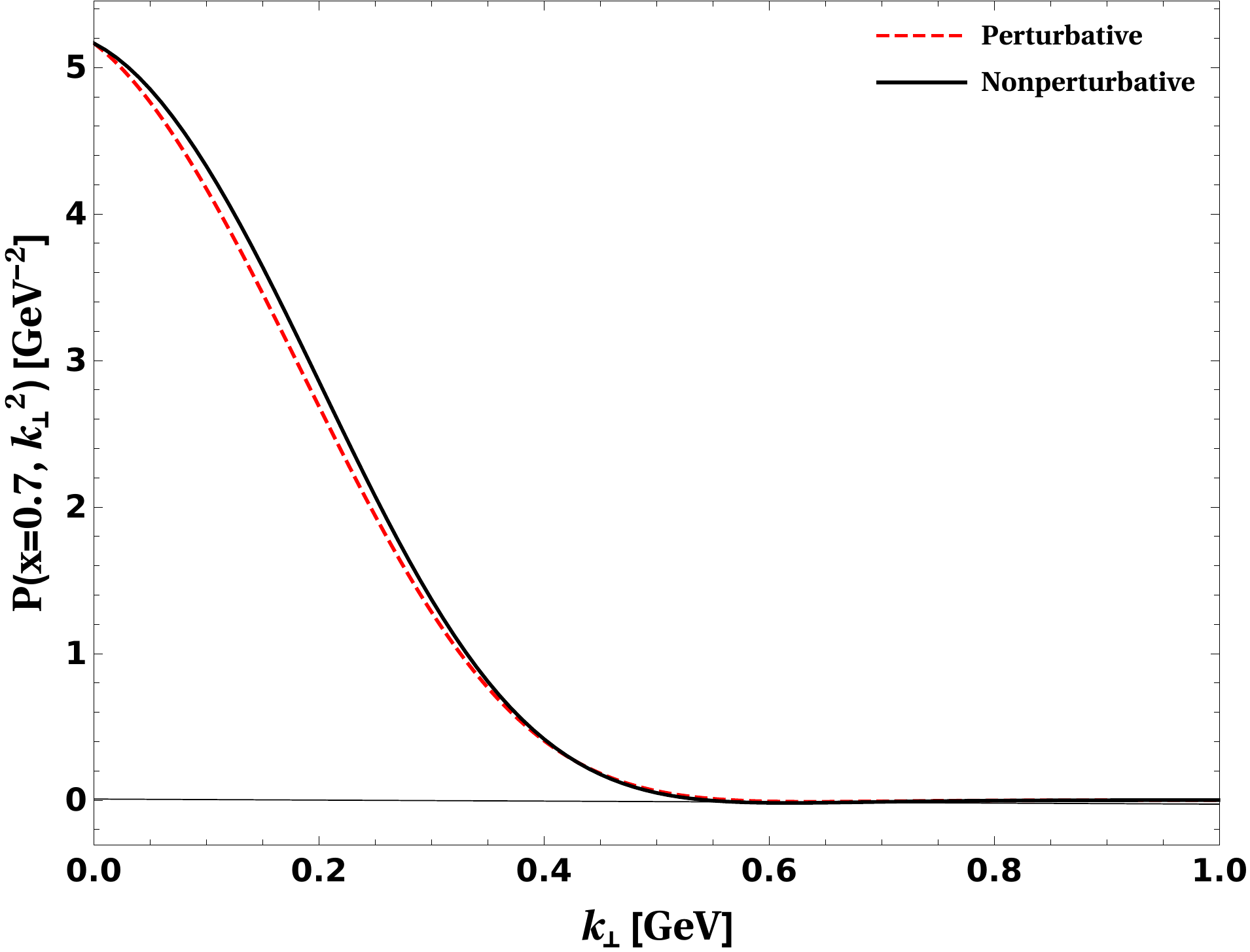}   
\caption{$P(x, k_\perp)$ at $x=0.1$ (upper left), $x=0.3$ (upper right), $x=0.5$ (lower left) and $x=0.7$ (lower right) with the holographic Boer-Mulders function generated by the perturbative kernel (red dashed curves) and the non-perturbative kernel (solid black curves).}
  \label{Fig:P}
  \end{figure}

 A model-independent theory constraint on our holographic TMDs is the positivity bound \cite{Bacchetta:1999kz}:
\begin{equation}
	P(x, k_\perp) \equiv f_{1}(x,k_\perp)-\frac{k_\perp}{M_\pi}|h_{1}^\perp(x, k_\perp)| \ge 0 \;.
\label{P}
\end{equation}
 In Fig. \ref{Fig:P}, we show that this constraint is safely satisfied when the holographic Boer-Mulders function is generated by the perturbative kernel with $\alpha_s=0.3$. This is not the case if we use $\alpha_s >0.3$, although the violation only occurs for large $k_\perp$. This is also the case when the holographic Boer-Mulders is generated by the non-perturbative kernel. As can be seen in Fig. \ref{Fig:P}, this violation becomes somewhat more pronounced (i.e. happening for smaller $k_\perp$) for small $x$. Similar violations of the positivity constraint have been reported in the literature \cite{Pasquini:2014ppa,Wang:2017onm,Kotzinian:2008fe}, albeit with the perturbative kernel. They seem to indicate a limitation of current non-perturbative models to accurately capture the large $k_\perp$ behaviour of the TMDs.    
 
 \section{Comparison to lattice}

 The generalized Boer-Mulders shifts of the pion TMDs have also been predicted from first principles in lattice QCD at a scale of $2$ GeV and using a pion mass, $M_\pi=518$ MeV  \cite{Engelhardt:2015xja}. In order to compare with the lattice predictions, we compute the shifts defined as
\begin{equation}
	\langle k_\perp \rangle_{UT} (b^2_\perp)= M_\pi \frac{\tilde{h}_{1,\pi}^{\perp [1](1)}(b_\perp^2)}{\tilde{f}_{1,\pi}^{[1](0)}(b_\perp^2)} \;,
\label{gBMshift}
\end{equation}
where the generalized TMD moments are given by
\begin{equation}
	\tilde{f}^{[m](n)}(b_\perp^2)=\frac{2\pi n !}{M_\pi^{2n}} \int \mathrm{d}x ~ x^{m-1} \int \mathrm{d} k_\perp ~ k_\perp \left(\frac{k_\perp}{b_\perp} \right)^n J_{n}(b_\perp k_\perp) f(x,k_\perp^2) \;.
\end{equation}
The $b_\perp \to 0$ limit of Eq. \eqref{gBMshift} is a measure of the quark's average transverse momentum in a direction perpendicular to its polarization. Our results are shown in Table \ref{Tab:gBMshift}. As can be seen, it is possible to fit the lattice data by using a large $\alpha_s$ with the perturbative kernel. However, as we mentioned earlier, $\alpha_s=0.9$ is perhaps not consistent with the weak coupling hypothesis $g^2 \ll 1$. We prefer to consider the predictions with $\alpha_s=0.3$ as a more realistic prediction with the perturbative kernel. Then, it becomes apparent that the non-perturbative kernel does a better job, bringing our predictions closer to the lattice data. It might be possible to improve upon these predictions by using a different non-perturbative gluon propagator than the one given by Eq. \eqref{DS-gluon} or a different set of fitted parameters in Eq. \eqref{DS-gluon} itself. Thus the non-perturbative kernel offers a more promising way to fit the lattice data.

We should also emphasize that our predictions are at a low hadronic scale and are obtained using the physical pion mass while the lattice predictions are at a scale of $2$ GeV and are obtained using a pion mass of $518$ MeV. We have checked that our predictions do not change much if we use the larger pion mass. On the other hand, we have not attempted to address here the more delicate issue  of evolving our holographic TMDs to a higher scale. Indeed, the evolution of TMDs are likely to be driven both by perturbative and non-perturbative physics \cite{Collins:2014jpa,Scimemi:2016ffw,DAlesio:2014mrz,Collins:1984kg} and are not yet fully known \cite{Rogers:2015sqa}. The approximate evolution of the original holographic unpolarized TMD up to $5$ GeV, has been carried out in Ref. \cite{Bacchetta:2017vzh}, revealing a substantial change in its width as well as the $x$-dependence of the latter.

 \begin{table}[hbt]
  \begin{tabular}{|l|c|c|c|c|c|}
  \hline
   $b_\perp$ & Lattice 1   &  Lattice 2  &  Non Pert. & Pert. [$\alpha_s=0.3$] & Pert. [$\alpha_s=0.9$]\\
    \hline
   0.27 & -0.138(28) & -0.133(19) & -0.0663 &-0.0424&-0.1273\\
   \hline
    0.34 & -0.128(29) & -0.121(16) & -0.0645 &-0.0423&-0.1268\\
    \hline
    0.36 & -0.145(25) & -0.148(15) &-0.0639  &-0.0422&-0.1267\\
   
    \hline
    
  \end{tabular}
  \caption{Our predictions, in GeV, for the generalized Boer-Mulders shifts given by Eq. \eqref{gBMshift} at different values of $b_\perp$ in fm. The two sets of lattice data are from Ref. \cite{Engelhardt:2015xja}. Our predictions are computed using the non-perturbatively and perturbatively generated holographic Boer-Mulders functions. The ``perturbative" predictions are given at two values of $\alpha_s$.}
  \label{Tab:gBMshift}
\end{table}

 \section{Conclusions}
 
 We have predicted the two leading twist pion TMDs using the spin-improved holographic light-front wavefunction for the pion. To predict the holographic Boer-Mulders function, we used both a perturbative and a non-perturbative gluon rescattering kernel.  We find that the non-perturbative kernel offers a more promising way to describe the available lattice data on the generalized Boer-Mulders shifts. A more precise comparison to the lattice data may be possible when the evolution of our holographic pion TMDs are taken into account and if lattice data become available at the physical pion mass.

   \section{Acknowledgements}
M.A and R.S are supported by Individual Discovery Grants from the Natural Science and Engineering Research Council of Canada (NSERC): SAPIN-2017-00033 and SAPIN-2017-00031 respectively. C.M is supported by  the Natural Science Foundation of China (NSFC) under the grant No. 11850410436. We thank L. Gamberg and M. Schlegel for useful discussions.
\bibliographystyle{apsrev}
\bibliography{PionTMD.bib}

\begin{thebibliography}{48}
\expandafter\ifx\csname natexlab\endcsname\relax\def\natexlab#1{#1}\fi
\expandafter\ifx\csname bibnamefont\endcsname\relax
  \def\bibnamefont#1{#1}\fi
\expandafter\ifx\csname bibfnamefont\endcsname\relax
  \def\bibfnamefont#1{#1}\fi
\expandafter\ifx\csname citenamefont\endcsname\relax
  \def\citenamefont#1{#1}\fi
\expandafter\ifx\csname url\endcsname\relax
  \def\url#1{\texttt{#1}}\fi
\expandafter\ifx\csname urlprefix\endcsname\relax\def\urlprefix{URL }\fi
\providecommand{\bibinfo}[2]{#2}
\providecommand{\eprint}[2][]{\url{#2}}

\bibitem[{\citenamefont{Angeles-Martinez
  et~al.}(2015)}]{Angeles-Martinez:2015sea}
\bibinfo{author}{\bibfnamefont{R.}~\bibnamefont{Angeles-Martinez}}
  \bibnamefont{et~al.}, \bibinfo{journal}{Acta Phys. Polon.}
  \textbf{\bibinfo{volume}{B46}}, \bibinfo{pages}{2501} (\bibinfo{year}{2015}),
  \eprint{1507.05267}.

\bibitem[{\citenamefont{Boer and Mulders}(1998)}]{Boer:1997nt}
\bibinfo{author}{\bibfnamefont{D.}~\bibnamefont{Boer}} \bibnamefont{and}
  \bibinfo{author}{\bibfnamefont{P.~J.} \bibnamefont{Mulders}},
  \bibinfo{journal}{Phys. Rev.} \textbf{\bibinfo{volume}{D57}},
  \bibinfo{pages}{5780} (\bibinfo{year}{1998}), \eprint{hep-ph/9711485}.

\bibitem[{\citenamefont{Boer}(1999)}]{Boer:1999mm}
\bibinfo{author}{\bibfnamefont{D.}~\bibnamefont{Boer}}, \bibinfo{journal}{Phys.
  Rev.} \textbf{\bibinfo{volume}{D60}}, \bibinfo{pages}{014012}
  (\bibinfo{year}{1999}), \eprint{hep-ph/9902255}.

\bibitem[{\citenamefont{Collins}(1993)}]{Collins:1992kk}
\bibinfo{author}{\bibfnamefont{J.~C.} \bibnamefont{Collins}},
  \bibinfo{journal}{Nucl. Phys.} \textbf{\bibinfo{volume}{B396}},
  \bibinfo{pages}{161} (\bibinfo{year}{1993}), \eprint{hep-ph/9208213}.

\bibitem[{\citenamefont{Brodsky
  et~al.}(2002{\natexlab{a}})\citenamefont{Brodsky, Hwang, and
  Schmidt}}]{Brodsky:2002cx}
\bibinfo{author}{\bibfnamefont{S.~J.} \bibnamefont{Brodsky}},
  \bibinfo{author}{\bibfnamefont{D.~S.} \bibnamefont{Hwang}}, \bibnamefont{and}
  \bibinfo{author}{\bibfnamefont{I.}~\bibnamefont{Schmidt}},
  \bibinfo{journal}{Phys. Lett.} \textbf{\bibinfo{volume}{B530}},
  \bibinfo{pages}{99} (\bibinfo{year}{2002}{\natexlab{a}}),
  \eprint{hep-ph/0201296}.

\bibitem[{\citenamefont{Brodsky
  et~al.}(2002{\natexlab{b}})\citenamefont{Brodsky, Hwang, and
  Schmidt}}]{Brodsky:2002rv}
\bibinfo{author}{\bibfnamefont{S.~J.} \bibnamefont{Brodsky}},
  \bibinfo{author}{\bibfnamefont{D.~S.} \bibnamefont{Hwang}}, \bibnamefont{and}
  \bibinfo{author}{\bibfnamefont{I.}~\bibnamefont{Schmidt}},
  \bibinfo{journal}{Nucl. Phys.} \textbf{\bibinfo{volume}{B642}},
  \bibinfo{pages}{344} (\bibinfo{year}{2002}{\natexlab{b}}),
  \eprint{hep-ph/0206259}.

\bibitem[{\citenamefont{Collins}(2002)}]{Collins:2002kn}
\bibinfo{author}{\bibfnamefont{J.~C.} \bibnamefont{Collins}},
  \bibinfo{journal}{Phys. Lett.} \textbf{\bibinfo{volume}{B536}},
  \bibinfo{pages}{43} (\bibinfo{year}{2002}), \eprint{hep-ph/0204004}.

\bibitem[{\citenamefont{Ji and Yuan}(2002)}]{Ji:2002aa}
\bibinfo{author}{\bibfnamefont{X.-d.} \bibnamefont{Ji}} \bibnamefont{and}
  \bibinfo{author}{\bibfnamefont{F.}~\bibnamefont{Yuan}},
  \bibinfo{journal}{Phys. Lett.} \textbf{\bibinfo{volume}{B543}},
  \bibinfo{pages}{66} (\bibinfo{year}{2002}), \eprint{hep-ph/0206057}.

\bibitem[{\citenamefont{Belitsky et~al.}(2003)\citenamefont{Belitsky, Ji, and
  Yuan}}]{Belitsky:2002sm}
\bibinfo{author}{\bibfnamefont{A.~V.} \bibnamefont{Belitsky}},
  \bibinfo{author}{\bibfnamefont{X.}~\bibnamefont{Ji}}, \bibnamefont{and}
  \bibinfo{author}{\bibfnamefont{F.}~\bibnamefont{Yuan}},
  \bibinfo{journal}{Nucl. Phys.} \textbf{\bibinfo{volume}{B656}},
  \bibinfo{pages}{165} (\bibinfo{year}{2003}), \eprint{hep-ph/0208038}.

\bibitem[{\citenamefont{Pasquini and Schweitzer}(2014)}]{Pasquini:2014ppa}
\bibinfo{author}{\bibfnamefont{B.}~\bibnamefont{Pasquini}} \bibnamefont{and}
  \bibinfo{author}{\bibfnamefont{P.}~\bibnamefont{Schweitzer}},
  \bibinfo{journal}{Phys. Rev.} \textbf{\bibinfo{volume}{D90}},
  \bibinfo{pages}{014050} (\bibinfo{year}{2014}), \eprint{1406.2056}.

\bibitem[{\citenamefont{Wang et~al.}(2017)\citenamefont{Wang, Wang, and
  Lu}}]{Wang:2017onm}
\bibinfo{author}{\bibfnamefont{Z.}~\bibnamefont{Wang}},
  \bibinfo{author}{\bibfnamefont{X.}~\bibnamefont{Wang}}, \bibnamefont{and}
  \bibinfo{author}{\bibfnamefont{Z.}~\bibnamefont{Lu}}, \bibinfo{journal}{Phys.
  Rev.} \textbf{\bibinfo{volume}{D95}}, \bibinfo{pages}{094004}
  (\bibinfo{year}{2017}), \eprint{1702.03637}.

\bibitem[{\citenamefont{Guanziroli et~al.}(1988)}]{Guanziroli:1987rp}
\bibinfo{author}{\bibfnamefont{M.}~\bibnamefont{Guanziroli}}
  \bibnamefont{et~al.} (\bibinfo{collaboration}{NA10}), \bibinfo{journal}{Z.
  Phys.} \textbf{\bibinfo{volume}{C37}}, \bibinfo{pages}{545}
  (\bibinfo{year}{1988}).

\bibitem[{\citenamefont{Falciano et~al.}(1986)}]{Falciano:1986wk}
\bibinfo{author}{\bibfnamefont{S.}~\bibnamefont{Falciano}} \bibnamefont{et~al.}
  (\bibinfo{collaboration}{NA10}), \bibinfo{journal}{Z. Phys.}
  \textbf{\bibinfo{volume}{C31}}, \bibinfo{pages}{513} (\bibinfo{year}{1986}).

\bibitem[{\citenamefont{Conway et~al.}(1989)}]{Conway:1989fs}
\bibinfo{author}{\bibfnamefont{J.~S.} \bibnamefont{Conway}}
  \bibnamefont{et~al.}, \bibinfo{journal}{Phys. Rev.}
  \textbf{\bibinfo{volume}{D39}}, \bibinfo{pages}{92} (\bibinfo{year}{1989}).

\bibitem[{\citenamefont{Aghasyan et~al.}(2017)}]{Aghasyan:2017jop}
\bibinfo{author}{\bibfnamefont{M.}~\bibnamefont{Aghasyan}} \bibnamefont{et~al.}
  (\bibinfo{collaboration}{COMPASS}), \bibinfo{journal}{Phys. Rev. Lett.}
  \textbf{\bibinfo{volume}{119}}, \bibinfo{pages}{112002}
  (\bibinfo{year}{2017}), \eprint{1704.00488}.

\bibitem[{\citenamefont{Gautheron et~al.}(2010)}]{COMPASS}
\bibinfo{author}{\bibfnamefont{F.}~\bibnamefont{Gautheron}}
  \bibnamefont{et~al.} (\bibinfo{collaboration}{COMPASS collaboration}),
  \bibinfo{journal}{SPSC-P-340 CERN-SPSC-2010-014}  (\bibinfo{year}{2010}).

\bibitem[{\citenamefont{Lu and Ma}(2004{\natexlab{a}})}]{Lu:2004au}
\bibinfo{author}{\bibfnamefont{Z.}~\bibnamefont{Lu}} \bibnamefont{and}
  \bibinfo{author}{\bibfnamefont{B.-Q.} \bibnamefont{Ma}},
  \bibinfo{journal}{Nucl. Phys.} \textbf{\bibinfo{volume}{A741}},
  \bibinfo{pages}{200} (\bibinfo{year}{2004}{\natexlab{a}}),
  \eprint{hep-ph/0406171}.

\bibitem[{\citenamefont{Meissner et~al.}(2008)\citenamefont{Meissner, Metz,
  Schlegel, and Goeke}}]{Meissner:2008ay}
\bibinfo{author}{\bibfnamefont{S.}~\bibnamefont{Meissner}},
  \bibinfo{author}{\bibfnamefont{A.}~\bibnamefont{Metz}},
  \bibinfo{author}{\bibfnamefont{M.}~\bibnamefont{Schlegel}}, \bibnamefont{and}
  \bibinfo{author}{\bibfnamefont{K.}~\bibnamefont{Goeke}},
  \bibinfo{journal}{JHEP} \textbf{\bibinfo{volume}{08}}, \bibinfo{pages}{038}
  (\bibinfo{year}{2008}), \eprint{0805.3165}.

\bibitem[{\citenamefont{Wang et~al.}(2018)\citenamefont{Wang, Mao, and
  Lu}}]{Wang:2018naw}
\bibinfo{author}{\bibfnamefont{X.}~\bibnamefont{Wang}},
  \bibinfo{author}{\bibfnamefont{W.}~\bibnamefont{Mao}}, \bibnamefont{and}
  \bibinfo{author}{\bibfnamefont{Z.}~\bibnamefont{Lu}}, \bibinfo{journal}{Eur.
  Phys. J.} \textbf{\bibinfo{volume}{C78}}, \bibinfo{pages}{643}
  (\bibinfo{year}{2018}), \eprint{1805.03017}.

\bibitem[{\citenamefont{Lorc\'e et~al.}(2016)\citenamefont{Lorc\'e, Pasquini,
  and Schweitzer}}]{Lorce:2016ugb}
\bibinfo{author}{\bibfnamefont{C.}~\bibnamefont{Lorc\'e}},
  \bibinfo{author}{\bibfnamefont{B.}~\bibnamefont{Pasquini}}, \bibnamefont{and}
  \bibinfo{author}{\bibfnamefont{P.}~\bibnamefont{Schweitzer}},
  \bibinfo{journal}{Eur. Phys. J.} \textbf{\bibinfo{volume}{C76}},
  \bibinfo{pages}{415} (\bibinfo{year}{2016}), \eprint{1605.00815}.

\bibitem[{\citenamefont{Lu et~al.}(2012)\citenamefont{Lu, Ma, and
  Zhu}}]{Lu:2012hh}
\bibinfo{author}{\bibfnamefont{Z.}~\bibnamefont{Lu}},
  \bibinfo{author}{\bibfnamefont{B.-Q.} \bibnamefont{Ma}}, \bibnamefont{and}
  \bibinfo{author}{\bibfnamefont{J.}~\bibnamefont{Zhu}},
  \bibinfo{journal}{Phys. Rev.} \textbf{\bibinfo{volume}{D86}},
  \bibinfo{pages}{094023} (\bibinfo{year}{2012}), \eprint{1211.1745}.

\bibitem[{\citenamefont{Noguera and Scopetta}(2015)}]{Noguera:2015iia}
\bibinfo{author}{\bibfnamefont{S.}~\bibnamefont{Noguera}} \bibnamefont{and}
  \bibinfo{author}{\bibfnamefont{S.}~\bibnamefont{Scopetta}},
  \bibinfo{journal}{JHEP} \textbf{\bibinfo{volume}{11}}, \bibinfo{pages}{102}
  (\bibinfo{year}{2015}), \eprint{1508.01061}.

\bibitem[{\citenamefont{Ceccopieri et~al.}(2018)\citenamefont{Ceccopieri,
  Courtoy, Noguera, and Scopetta}}]{Ceccopieri:2018nop}
\bibinfo{author}{\bibfnamefont{F.~A.} \bibnamefont{Ceccopieri}},
  \bibinfo{author}{\bibfnamefont{A.}~\bibnamefont{Courtoy}},
  \bibinfo{author}{\bibfnamefont{S.}~\bibnamefont{Noguera}}, \bibnamefont{and}
  \bibinfo{author}{\bibfnamefont{S.}~\bibnamefont{Scopetta}},
  \bibinfo{journal}{Eur. Phys. J.} \textbf{\bibinfo{volume}{C78}},
  \bibinfo{pages}{644} (\bibinfo{year}{2018}), \eprint{1801.07682}.

\bibitem[{\citenamefont{Gamberg and Schlegel}(2010)}]{Gamberg:2009uk}
\bibinfo{author}{\bibfnamefont{L.}~\bibnamefont{Gamberg}} \bibnamefont{and}
  \bibinfo{author}{\bibfnamefont{M.}~\bibnamefont{Schlegel}},
  \bibinfo{journal}{Phys. Lett.} \textbf{\bibinfo{volume}{B685}},
  \bibinfo{pages}{95} (\bibinfo{year}{2010}), \eprint{0911.1964}.

\bibitem[{\citenamefont{Brommel et~al.}(2008)}]{Brommel:2007xd}
\bibinfo{author}{\bibfnamefont{D.}~\bibnamefont{Brommel}} \bibnamefont{et~al.}
  (\bibinfo{collaboration}{QCDSF, UKQCD}), \bibinfo{journal}{Phys. Rev. Lett.}
  \textbf{\bibinfo{volume}{101}}, \bibinfo{pages}{122001}
  (\bibinfo{year}{2008}), \eprint{0708.2249}.

\bibitem[{\citenamefont{Engelhardt et~al.}(2016)\citenamefont{Engelhardt,
  Hagler, Musch, Negele, and Schafer}}]{Engelhardt:2015xja}
\bibinfo{author}{\bibfnamefont{M.}~\bibnamefont{Engelhardt}},
  \bibinfo{author}{\bibfnamefont{P.}~\bibnamefont{Hagler}},
  \bibinfo{author}{\bibfnamefont{B.}~\bibnamefont{Musch}},
  \bibinfo{author}{\bibfnamefont{J.}~\bibnamefont{Negele}}, \bibnamefont{and}
  \bibinfo{author}{\bibfnamefont{A.}~\bibnamefont{Schafer}},
  \bibinfo{journal}{Phys. Rev.} \textbf{\bibinfo{volume}{D93}},
  \bibinfo{pages}{054501} (\bibinfo{year}{2016}), \eprint{1506.07826}.

\bibitem[{\citenamefont{Ahmady et~al.}(2017)\citenamefont{Ahmady, Chishtie, and
  Sandapen}}]{Ahmady:2016ufq}
\bibinfo{author}{\bibfnamefont{M.}~\bibnamefont{Ahmady}},
  \bibinfo{author}{\bibfnamefont{F.}~\bibnamefont{Chishtie}}, \bibnamefont{and}
  \bibinfo{author}{\bibfnamefont{R.}~\bibnamefont{Sandapen}},
  \bibinfo{journal}{Phys. Rev.} \textbf{\bibinfo{volume}{D95}},
  \bibinfo{pages}{074008} (\bibinfo{year}{2017}), \eprint{1609.07024}.

\bibitem[{\citenamefont{Ahmady et~al.}(2018)\citenamefont{Ahmady, Mondal, and
  Sandapen}}]{Ahmady:2018muv}
\bibinfo{author}{\bibfnamefont{M.}~\bibnamefont{Ahmady}},
  \bibinfo{author}{\bibfnamefont{C.}~\bibnamefont{Mondal}}, \bibnamefont{and}
  \bibinfo{author}{\bibfnamefont{R.}~\bibnamefont{Sandapen}},
  \bibinfo{journal}{Phys. Rev.} \textbf{\bibinfo{volume}{D98}},
  \bibinfo{pages}{034010} (\bibinfo{year}{2018}), \eprint{1805.08911}.

\bibitem[{\citenamefont{de~T\'eramond and Brodsky}(2005)}]{deTeramond:2005su}
\bibinfo{author}{\bibfnamefont{G.~F.} \bibnamefont{de~T\'eramond}}
  \bibnamefont{and} \bibinfo{author}{\bibfnamefont{S.~J.}
  \bibnamefont{Brodsky}}, \bibinfo{journal}{Phys. Rev. Lett.}
  \textbf{\bibinfo{volume}{94}}, \bibinfo{pages}{201601}
  (\bibinfo{year}{2005}), \eprint{hep-th/0501022}.

\bibitem[{\citenamefont{Brodsky and de~T\'eramond}(2006)}]{Brodsky:2006uqa}
\bibinfo{author}{\bibfnamefont{S.~J.} \bibnamefont{Brodsky}} \bibnamefont{and}
  \bibinfo{author}{\bibfnamefont{G.~F.} \bibnamefont{de~T\'eramond}},
  \bibinfo{journal}{Phys. Rev. Lett.} \textbf{\bibinfo{volume}{96}},
  \bibinfo{pages}{201601} (\bibinfo{year}{2006}), \eprint{hep-ph/0602252}.

\bibitem[{\citenamefont{de~T\'eramond and Brodsky}(2009)}]{deTeramond:2008ht}
\bibinfo{author}{\bibfnamefont{G.~F.} \bibnamefont{de~T\'eramond}}
  \bibnamefont{and} \bibinfo{author}{\bibfnamefont{S.~J.}
  \bibnamefont{Brodsky}}, \bibinfo{journal}{Phys. Rev. Lett.}
  \textbf{\bibinfo{volume}{102}}, \bibinfo{pages}{081601}
  (\bibinfo{year}{2009}), \eprint{0809.4899}.

\bibitem[{\citenamefont{de~Alfaro et~al.}(1976)\citenamefont{de~Alfaro, Fubini,
  and Furlan}}]{deAlfaro:1976vlx}
\bibinfo{author}{\bibfnamefont{V.}~\bibnamefont{de~Alfaro}},
  \bibinfo{author}{\bibfnamefont{S.}~\bibnamefont{Fubini}}, \bibnamefont{and}
  \bibinfo{author}{\bibfnamefont{G.}~\bibnamefont{Furlan}},
  \bibinfo{journal}{Nuovo Cim.} \textbf{\bibinfo{volume}{A34}},
  \bibinfo{pages}{569} (\bibinfo{year}{1976}).

\bibitem[{\citenamefont{Brodsky et~al.}(2016)\citenamefont{Brodsky,
  de~T\'eramond, Dosch, and Lorce}}]{Brodsky:2016rvj}
\bibinfo{author}{\bibfnamefont{S.~J.} \bibnamefont{Brodsky}},
  \bibinfo{author}{\bibfnamefont{G.~F.} \bibnamefont{de~T\'eramond}},
  \bibinfo{author}{\bibfnamefont{H.~G.} \bibnamefont{Dosch}}, \bibnamefont{and}
  \bibinfo{author}{\bibfnamefont{C.}~\bibnamefont{Lorce}},
  \bibinfo{journal}{Int. J. Mod. Phys.} \textbf{\bibinfo{volume}{A31}},
  \bibinfo{pages}{1630029} (\bibinfo{year}{2016}), \eprint{1606.04638}.

\bibitem[{\citenamefont{Brodsky et~al.}(2015)\citenamefont{Brodsky,
  de~T\'eramond, Dosch, and Erlich}}]{Brodsky:2014yha}
\bibinfo{author}{\bibfnamefont{S.~J.} \bibnamefont{Brodsky}},
  \bibinfo{author}{\bibfnamefont{G.~F.} \bibnamefont{de~T\'eramond}},
  \bibinfo{author}{\bibfnamefont{H.~G.} \bibnamefont{Dosch}}, \bibnamefont{and}
  \bibinfo{author}{\bibfnamefont{J.}~\bibnamefont{Erlich}},
  \bibinfo{journal}{Phys. Rept.} \textbf{\bibinfo{volume}{584}},
  \bibinfo{pages}{1} (\bibinfo{year}{2015}), \eprint{1407.8131}.

\bibitem[{\citenamefont{Brodsky and de~T\'eramond}(2008)}]{Brodsky:2008pf}
\bibinfo{author}{\bibfnamefont{S.~J.} \bibnamefont{Brodsky}} \bibnamefont{and}
  \bibinfo{author}{\bibfnamefont{G.~F.} \bibnamefont{de~T\'eramond}},
  \bibinfo{journal}{Phys. Rev.} \textbf{\bibinfo{volume}{D78}},
  \bibinfo{pages}{025032} (\bibinfo{year}{2008}), \eprint{0804.0452}.

\bibitem[{\citenamefont{Bacchetta et~al.}(2017)\citenamefont{Bacchetta,
  Cotogno, and Pasquini}}]{Bacchetta:2017vzh}
\bibinfo{author}{\bibfnamefont{A.}~\bibnamefont{Bacchetta}},
  \bibinfo{author}{\bibfnamefont{S.}~\bibnamefont{Cotogno}}, \bibnamefont{and}
  \bibinfo{author}{\bibfnamefont{B.}~\bibnamefont{Pasquini}},
  \bibinfo{journal}{Phys. Lett.} \textbf{\bibinfo{volume}{B771}},
  \bibinfo{pages}{546} (\bibinfo{year}{2017}), \eprint{1703.07669}.

\bibitem[{\citenamefont{Lepage and Brodsky}(1980)}]{Lepage:1980fj}
\bibinfo{author}{\bibfnamefont{G.~P.} \bibnamefont{Lepage}} \bibnamefont{and}
  \bibinfo{author}{\bibfnamefont{S.~J.} \bibnamefont{Brodsky}},
  \bibinfo{journal}{Phys. Rev.} \textbf{\bibinfo{volume}{D22}},
  \bibinfo{pages}{2157} (\bibinfo{year}{1980}).

\bibitem[{\citenamefont{Aicher et~al.}(2010)\citenamefont{Aicher, Schafer, and
  Vogelsang}}]{Aicher:2010cb}
\bibinfo{author}{\bibfnamefont{M.}~\bibnamefont{Aicher}},
  \bibinfo{author}{\bibfnamefont{A.}~\bibnamefont{Schafer}}, \bibnamefont{and}
  \bibinfo{author}{\bibfnamefont{W.}~\bibnamefont{Vogelsang}},
  \bibinfo{journal}{Phys. Rev. Lett.} \textbf{\bibinfo{volume}{105}},
  \bibinfo{pages}{252003} (\bibinfo{year}{2010}), \eprint{1009.2481}.

\bibitem[{\citenamefont{Lu and Ma}(2004{\natexlab{b}})}]{Lu:2004hu}
\bibinfo{author}{\bibfnamefont{Z.}~\bibnamefont{Lu}} \bibnamefont{and}
  \bibinfo{author}{\bibfnamefont{B.-Q.} \bibnamefont{Ma}},
  \bibinfo{journal}{Phys. Rev.} \textbf{\bibinfo{volume}{D70}},
  \bibinfo{pages}{094044} (\bibinfo{year}{2004}{\natexlab{b}}),
  \eprint{hep-ph/0411043}.

\bibitem[{\citenamefont{Burkardt and Hannafious}(2008)}]{Burkardt:2007xm}
\bibinfo{author}{\bibfnamefont{M.}~\bibnamefont{Burkardt}} \bibnamefont{and}
  \bibinfo{author}{\bibfnamefont{B.}~\bibnamefont{Hannafious}},
  \bibinfo{journal}{Phys. Lett.} \textbf{\bibinfo{volume}{B658}},
  \bibinfo{pages}{130} (\bibinfo{year}{2008}), \eprint{0705.1573}.

\bibitem[{\citenamefont{Fischer and Alkofer}(2003)}]{Fischer:2003rp}
\bibinfo{author}{\bibfnamefont{C.~S.} \bibnamefont{Fischer}} \bibnamefont{and}
  \bibinfo{author}{\bibfnamefont{R.}~\bibnamefont{Alkofer}},
  \bibinfo{journal}{Phys. Rev.} \textbf{\bibinfo{volume}{D67}},
  \bibinfo{pages}{094020} (\bibinfo{year}{2003}), \eprint{hep-ph/0301094}.

\bibitem[{\citenamefont{Bacchetta et~al.}(2000)\citenamefont{Bacchetta,
  Boglione, Henneman, and Mulders}}]{Bacchetta:1999kz}
\bibinfo{author}{\bibfnamefont{A.}~\bibnamefont{Bacchetta}},
  \bibinfo{author}{\bibfnamefont{M.}~\bibnamefont{Boglione}},
  \bibinfo{author}{\bibfnamefont{A.}~\bibnamefont{Henneman}}, \bibnamefont{and}
  \bibinfo{author}{\bibfnamefont{P.~J.} \bibnamefont{Mulders}},
  \bibinfo{journal}{Phys. Rev. Lett.} \textbf{\bibinfo{volume}{85}},
  \bibinfo{pages}{712} (\bibinfo{year}{2000}), \eprint{hep-ph/9912490}.

\bibitem[{\citenamefont{Kotzinian}(2008)}]{Kotzinian:2008fe}
\bibinfo{author}{\bibfnamefont{A.}~\bibnamefont{Kotzinian}}, in
  \emph{\bibinfo{booktitle}{{Transversity 2008: 2nd International Workshop on
  Transverse Polarization Phenomena in Hard Processes Ferrara, Italy, May
  28-31, 2008}}} (\bibinfo{year}{2008}), \eprint{0806.3804}.

\bibitem[{\citenamefont{Collins and Rogers}(2015)}]{Collins:2014jpa}
\bibinfo{author}{\bibfnamefont{J.}~\bibnamefont{Collins}} \bibnamefont{and}
  \bibinfo{author}{\bibfnamefont{T.}~\bibnamefont{Rogers}},
  \bibinfo{journal}{Phys. Rev.} \textbf{\bibinfo{volume}{D91}},
  \bibinfo{pages}{074020} (\bibinfo{year}{2015}), \eprint{1412.3820}.

\bibitem[{\citenamefont{Scimemi and Vladimirov}(2017)}]{Scimemi:2016ffw}
\bibinfo{author}{\bibfnamefont{I.}~\bibnamefont{Scimemi}} \bibnamefont{and}
  \bibinfo{author}{\bibfnamefont{A.}~\bibnamefont{Vladimirov}},
  \bibinfo{journal}{JHEP} \textbf{\bibinfo{volume}{03}}, \bibinfo{pages}{002}
  (\bibinfo{year}{2017}), \eprint{1609.06047}.

\bibitem[{\citenamefont{D'Alesio et~al.}(2014)\citenamefont{D'Alesio,
  Echevarria, Melis, and Scimemi}}]{DAlesio:2014mrz}
\bibinfo{author}{\bibfnamefont{U.}~\bibnamefont{D'Alesio}},
  \bibinfo{author}{\bibfnamefont{M.~G.} \bibnamefont{Echevarria}},
  \bibinfo{author}{\bibfnamefont{S.}~\bibnamefont{Melis}}, \bibnamefont{and}
  \bibinfo{author}{\bibfnamefont{I.}~\bibnamefont{Scimemi}},
  \bibinfo{journal}{JHEP} \textbf{\bibinfo{volume}{11}}, \bibinfo{pages}{098}
  (\bibinfo{year}{2014}), \eprint{1407.3311}.

\bibitem[{\citenamefont{Collins et~al.}(1985)\citenamefont{Collins, Soper, and
  Sterman}}]{Collins:1984kg}
\bibinfo{author}{\bibfnamefont{J.~C.} \bibnamefont{Collins}},
  \bibinfo{author}{\bibfnamefont{D.~E.} \bibnamefont{Soper}}, \bibnamefont{and}
  \bibinfo{author}{\bibfnamefont{G.~F.} \bibnamefont{Sterman}},
  \bibinfo{journal}{Nucl. Phys.} \textbf{\bibinfo{volume}{B250}},
  \bibinfo{pages}{199} (\bibinfo{year}{1985}).

\bibitem[{\citenamefont{Rogers}(2016)}]{Rogers:2015sqa}
\bibinfo{author}{\bibfnamefont{T.~C.} \bibnamefont{Rogers}},
  \bibinfo{journal}{Eur. Phys. J.} \textbf{\bibinfo{volume}{A52}},
  \bibinfo{pages}{153} (\bibinfo{year}{2016}), \eprint{1509.04766}.

\end{thebibliography}

\end{document}